\documentclass[%
 reprint,
superscriptaddress,
 amsmath,amssymb,
 aps
]{revtex4-2}

\usepackage[T1]{fontenc}
\usepackage[utf8]{inputenc}
\usepackage{graphicx}
\usepackage{dcolumn}
\usepackage{bm}
\usepackage{xcolor}
\usepackage{multirow}
\usepackage{hhline}
\usepackage{makecell}

\usepackage{tabularx}
\usepackage{array}
\newcolumntype{Y}{>{\centering\arraybackslash}X}
\newcolumntype{C}[1]{>{\centering\arraybackslash}p{#1}}

\usepackage{float}
\usepackage{placeins}

\usepackage{comment}

\begin{document}

\preprint{APS/123-QED}

\title{\Large \textbf{Multivalley 3D Electronic Structure of PbSe from Soft-X-Ray ARPES and First-Principles Calculations}}

\author{Zefeng Cai}
\affiliation{Department of Materials Science and Engineering, Carnegie Mellon University, Pittsburgh, PA 15213, USA}
\author{Valentine V. Volobuev }
\affiliation{International Research Centre MagTop, Institute of Physics, Polish Academy of Sciences, al. Lotników 32/46, 02668 Warsaw, Poland}
\affiliation{National Technical University ``KhPI'', Kyrpychova Str. 2, 61002 Kharkiv, Ukraine}
\author{Jędrzej Korczak}
\affiliation{International Research Centre MagTop, Institute of Physics, Polish Academy of Sciences, al. Lotników 32/46, 02668 Warsaw, Poland}
\affiliation{Institute of Physics, Polish Academy of Sciences, al. Lotników 32/46, 02668 Warsaw, Poland}
\author{Enrico Della Valle}
\affiliation{Paul Scherrer Institut, Swiss Light Source, CH-5232 Villigen PSI, Switzerland}
\author{Hantian Liu}
\affiliation{Department of Materials Science and Engineering, Carnegie Mellon University, Pittsburgh, PA 15213, USA}
\author{Moritz Hoesch}
\affiliation{Deutsches Elektronen-Synchrotron DESY, Notkestra{\ss}e 85, 22607 Hamburg, Germany}
\author{Sergey M. Frolov}
\affiliation{Department of Physics and Astronomy, University of Pittsburgh, Pittsburgh, PA, 15260, USA} 
\author{Tomasz Story }
\email{Electronic mail: story@ifpan.edu.pl}
\affiliation{International Research Centre MagTop, Institute of Physics, Polish Academy of Sciences, al. Lotników 32/46, 02668 Warsaw, Poland}
\affiliation{Institute of Physics, Polish Academy of Sciences, al. Lotników 32/46, 02668 Warsaw, Poland}
\author{Vladimir N. Strocov}
\thanks{Electronic mail: vladimir.strocov@psi.ch}
\affiliation{Paul Scherrer Institut, Swiss Light Source, CH-5232 Villigen PSI, Switzerland}
\author{Noa Marom}
\email{Electronic mail: nmarom@andrew.cmu.edu}
 \affiliation{Department of Materials Science and Engineering, Carnegie Mellon University, Pittsburgh, PA 15213, USA}
\affiliation{Department of Chemistry, Carnegie Mellon University, Pittsburgh, PA 15213, USA}
\affiliation{Department of Physics, Carnegie Mellon University, Pittsburgh, PA 15213, USA}

\date{\today}

\begin{abstract}
PbSe is a narrow-gap IV-VI semiconductor, whose multivalley valence bands, with maxima at the L, $\Sigma$, and $\Delta$ points, underpin its intermediate-temperature thermoelectric properties. We combine soft-X-ray angle-resolved photoemission spectroscopy (SX-ARPES) with first principles simulations to study the valence band structure of bulk PbSe. High resolution measurements are conducted at photon energies of 400--900~eV to map the valence manifold along X$\Gamma$X, WXW, and K$\Gamma$K, and iso-energy surfaces are collected in the $k_z=0$ plane. Comparison to ARPES enables a rigorous assessment of the performance of density functional theory (DFT), using semi-local and hybrid functionals, as well as many-body perturbation theory within the quasiparticle self-consistent $GW$ approximation. We find that the Heyd-Scuseria-Ernzerhof (HSE) hybrid functional and QP$GW$ reproduce the measured band dispersions to within 0.1--0.2~eV over the entire valence band. In contrast, the semi-local Perdew-Burke-Ernzerhof (PBE) functional compresses the band width and deviates from experiment by up to 0.6~eV. We further show that an accurate band structure and band gap are vital to obtaining a correct description of the dependence of the Seebeck coefficient of p-type PbSe on the hole concentration (Pisarenko relation). This has implications for computational efforts to discover thermoelectric materials.

\end{abstract}

\maketitle

\section{\label{sec:level1} Introduction}

The IV-VI compounds PbTe, PbSe, PbS, and SnTe, as well as their substitutional solid solutions Pb$_{1-x}$Sn$_x$Te and Pb$_{1-x}$Sn$_x$Se ($x \leq 0.4$), are a family of narrow-gap semiconductors that crystallize in the rock-salt structure. They have a direct fundamental energy gap, which is useful for optoelectronic devices. PbSe has a gap of $E_g = 0.278$~eV at room temperature and $0.145$~eV at $T = 4$~K~\cite{Preier1979_PbSaltLasers}. They have a multivalley band structure with four conduction-band and valence-band valleys located at the L$\langle 111\rangle$ points of the first Brillouin zone~\cite{Mitchell1966_PbSe_kp,Dalven1973_PbSeBandGap}, which is beneficial for thermoelectricity. Thanks to these band structure features, these IV-VI compounds exhibit excellent thermoelectric and infrared optoelectronic properties, exploited in thermoelectric generators for heat-to-electricity conversion and in mid- and far-infrared detectors and junction lasers~\cite{Ravich1970_LeadChalcogenides,Khokhlov2002_LeadChalcogenides,Preier1979_PbSaltLasers,Wang2011_PbSe_thermoelectric,Qin2024_Science_gridplain}.
When alloyed with Sn, both the telluride and selenide materials exhibit intriguing effects, including electron-band inversion (topological crystalline insulators~\cite{Hsieh2012_TCI_SnTe,Dziawa2012_TCI_PbSnSe,AndoFu2015_TCI_review}), ferroelectric lattice distortion~\cite{Khokhlov2002_LeadChalcogenides,Wojek2015_DiracGap}, and carrier-induced ferromagnetism (in materials with magnetic Mn ions)~\cite{Eggenkamp1995_RKKY}.
Because they comprise heavy elements, they additionally exhibit strong relativistic effects (spin-orbit and Darwin terms), which determine the energy gaps, the ordering of the bands, and the parity of the states at the bottom of the conduction band and at the top of the valence band~\cite{Wei1997_PbSeBS,Hummer2007_PbChalcogenides_GW,Svane2010_QPGW_PbChalcogenides}. The  strong spin-orbit coupling also makes the IV-VI semiconductors attractive for topological quantum devices designed to host Majorana zero modes at semiconductor-superconductor interfaces~\cite{Frolov2020_TopoSC_hybrid}. Such devices were initially developed using III-V semiconductors~\cite{Pribiag2015_edgeSC,Jardine2025_InAsAl_barriers} and are being extended to the IV-VI family via the heteroepitaxial integration of PbSe films with III-V substrates~\cite{Haidet2023_PbSe_IIIV} and the growth of PbTe nanowires~\cite{Schellingerhout2022_PbTe_NW,Gupta2024_PbTe_JJ}.

\begin{figure}[!htbp]
\centering
\includegraphics[width=0.9\linewidth]{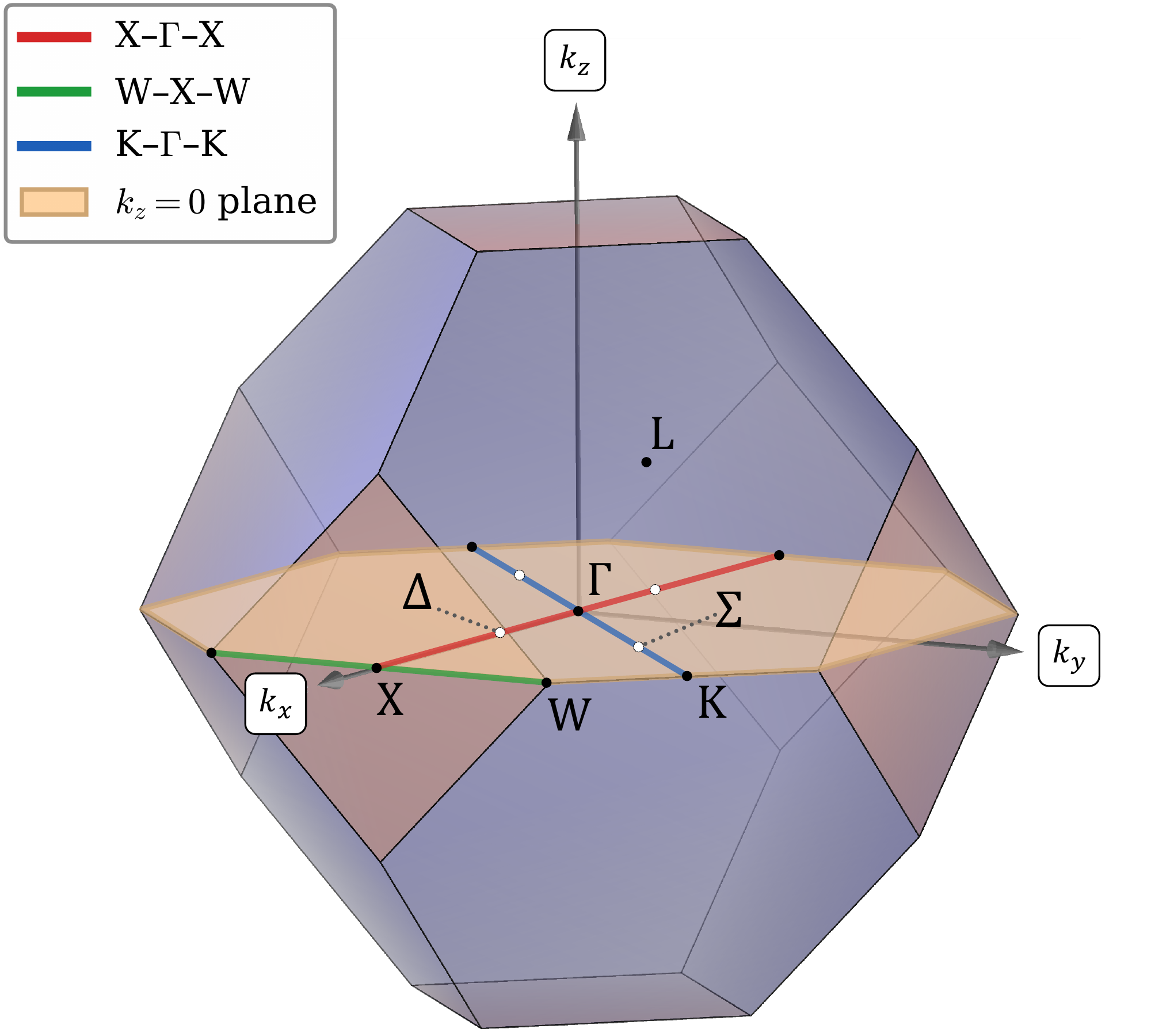}
\caption{First Brillouin zone of PbSe (rock-salt FCC), showing the three directions probed by ARPES in this work: X--$\Gamma$--X (red), W--X--W (green), and K--$\Gamma$--K (blue), all lying in the $k_z = 0$ plane (orange octagon), with $k_x$ and $k_y$ parallel and $k_z$ perpendicular to the (001) surface at which the ARPES spectra were measured.
The $\Sigma$ and $\Delta$ points, which are local maxima of the topmost valence band along $\Gamma$--K and $\Gamma$--X, respectively, are also annotated (white dots) on the corresponding paths.}
\label{fig:BZ}
\end{figure}

The current understanding of the electronic structure of IV-VI semiconductors is based on  optical~\cite{Dalven1973_PbSeBandGap}, magneto-transport~\cite{Ravich1970_LeadChalcogenides}, and thermoelectric studies~\cite{Pei2011_PbSeConvergence,Wang2011_PbSe_thermoelectric,Liu2025_PbSe_TE_review}, as well as on early valence-band photoemission experiments~\cite{McFeely1973_XPS_PbChalcogenides,Grandke1978_ARPES_PbChalcogenides,Hinkel1989_PbSe_bulkBands}.
The valence band edge is formed by the light-hole valleys at the L-point. A second valence band of heavy holes has a maximum located at the $\Sigma$~point along the $\Gamma$--K $\langle 110\rangle$ direction (Figure~\ref{fig:BZ}), with 12 equivalent valleys. This band also contributes significantly to the thermoelectric properties of $p$-doped materials~\cite{Wang2011_PbSe_thermoelectric,Chasapis2015_PbSe_twoband,Zhu2022_multiband}.
It has been observed by temperature-dependent ARPES that With increasing temperature the $\Sigma$ maximum and the L maximum approach each other~\cite{Zhao2017_ARPES_PbSe_convergence}.
A third maximum of the topmost valence band at the $\Delta$~point along the $\Gamma$--X $\langle 100\rangle$ direction was first suggested based on a semi-empirical model~\cite{Martinez1975_PbSe_EPM}. Subsequently, the position of the $\Delta$ maximum was probed by two  ARPES experiments, whose results disagreed  with each other by $\sim 0.5$~eV~\cite{Grandke1978_ARPES_PbChalcogenides,Hinkel1989_PbSe_bulkBands}. The relative energies of the three maxima at L, $\Sigma$, and $\Delta$ are important because simultaneous conduction through all three maxima would correspond to a record multivalency of 22 ($8 \times 1/2$ at L $+\, 12$ at $\Sigma$ $+\, 6$ at $\Delta$). Resolving these valleys experimentally requires a bulk-sensitive, \textbf{k}-space-resolved probe able to identify band positions with an energy precision of 0.1--0.2~eV.

On the computational side, there have been several first principles studies of PbSe and related materials. 
An early density functional theory (DFT) study employed the local density approximation (LDA) to study the electronic structure of the lead-chalcogenide series~\cite{Wei1997_PbSeBS}. However, the LDA gap of PbSe is not merely underestimated, but also inverted~\cite{Svane2010_QPGW_PbChalcogenides}. the generalized gradient approximation (GGA) of Perdew, Burke, and Ernzerhof (PBE)~\cite{1996_Perdew_GeneralizedGradientApproximation_PRL} likewise yields strongly underestimated, and even inverted, band gaps for all three lead chalcogenides~\cite{Hummer2007_PbChalcogenides_GW}. This is a manifestation of the self-interaction error (SIE), a spurious repulsion of an electron from its own charge density that arises in (semi-)local DFT functionals because the self-interaction in the Hartree term is not completely canceled out by the approximate exchange term~\cite{SIE,Cohen2012_DFT_challenges}. 
The screened hybrid functional of Heyd, Scuseria, and Ernzerhof (HSE) mitigates the SIE by replacing one quarter of the short-range semi-local exchange with exact (Fock) exchange~\cite{2003_Heyd_HSE_JChemPhys,2006_Krukau_HSE_JChemPhys,Paier2006_HSE_benchmark,Paier2006_HSE_erratum}. This brings the band gaps and effective masses of the lead chalcogenides into markedly better agreement with experiment~\cite{Hummer2007_PbChalcogenides_GW}.
The $GW$ approximation, within the framework of many-body perturbation theory, replaces the exchange-correlation potential of DFT altogether by an electron self-energy constructed from the one-electron Green's function $G$ and the screened Coulomb interaction $W$. The resulting effective potential is non-local and energy dependent, and the screening it contains is the computed dielectric response instead of a preset exchange fraction~\cite{Aryasetiawan1998_GW_RepProgPhys,Onida2002_GW_RMP}. In the quasiparticle self-consistent variant (QP$GW$), $G$ and $W$ are iterated to self-consistency, which removes the dependence on the DFT starting point~\cite{vanSchilfgaarde2006_QSGW,Shishkin2007_QPGW}. QP$GW$ with spin-orbit coupling restores the correct band ordering at the L-point for the lead chalcogenides~\cite{Svane2010_QPGW_PbChalcogenides}. However, QP$GW$ tends to overestimate band gaps as a consequence of neglecting the vertex~\cite{Svane2010_QPGW_PbChalcogenides,Shishkin2007_vertex,Grumet2018_QPGW_overestimate}. More recently, the temperature-induced renormalization of the band structure of PbTe has been computed by combining density functional perturbation theory with the Allen-Heine-Cardona theory of electron-phonon coupling~\cite{Querales2019_PbTe_Tbands}.
All of these methods have been validated almost exclusively against experimental band gaps and effective masses, which probe the bands only at their extrema. A rigorous benchmark of computed band structures against momentum-resolved experiments across the Brillouin zone has been lacking.

First-principles calculations of the thermoelectric transport coefficients of the IV-VI semiconductors have largely relied on semi-local DFT. A figure of merit as high as $zT \approx 2$ has been predicted for heavily doped p-type PbSe~\cite{Parker2010_PbSe_thermoelectric}. However, this was based on a GGA band structure with a gap of 65~meV, compared with the experimental value of $0.27$~eV. In an attempt to mitigate the band-gap underestimation of standard GGA functionals, another study of the doping- and temperature-dependent thermopower of PbTe employed the Engel-Vosko GGA~\cite{Singh2010_PbTe_thermopower}, which improves the band gaps at the expense of total energies and equilibrium lattice constants~\cite{Engel1993_EV_GGA,Dufek1994_EV_GGA_solids,Tran2017_KED_band_gaps}. A high-throughput screening study of the thermoelectric properties of materials from the Materials Project based on the PBE functional reported ``fair'' agreement with experiment for the maximum Seebeck coefficient across the dataset~\cite{Chen2016_CRTA_limits}. Therein, it was found that a rigid shift of the PBE conduction bands to reproduce the experimental band gaps improved the agreement with measured Seebeck coefficients. However, this is only possible if the band gap of a given material is known. Subsequently, it has been shown that using a hybrid functional~\cite{Berland2018_PbTe_hybrid_TE, Ganose2021_AMSET} or $GW$~\cite{Shi2015_SnSe_GW_TE} improves the agreement of thermoelectric properties with experimental data and increases the chances of discovering candidate materials that had been previously overlooked because PBE reduces their band gap to zero~\cite{2021_Berland_ApplPhysLett}.
More recent studies of the thermoelectric properties of chalcogenides have focused on going beyond the constant relaxation time approximation (CRTA; see Section~\ref{sec:te_formalism}) by considering electron-phonon coupling~\cite{Askarpour2023_PbX_intervalley, Askarpour2019_GeTe_anisotropy, Chaves2021_SnSe_scattering} and explicitly treating the carrier relaxation time~\cite{Xi2018_chalcogenide_screening, 2021_Fan_JMCC, Ganose2021_AMSET}, while keeping the exchange-correlation treatment at the semi-local level.

Here, we study PbSe as a representative example, which exhibits all the key valence-band features of the IV-VI semiconductor family. In contrast to PbTe, PbSe cleaves easily under ultrahigh-vacuum conditions, exposing an atomically flat and clean (001) crystal plane, thereby producing optimal samples for ARPES. This has facilitated ARPES studies of the valence band of Pb$_{1-x}$Sn$_x$Se ($x = 0$--$0.4$) topological crystals using photon energies of 10--40~eV~\cite{Dziawa2012_TCI_PbSnSe,Wojek2014_PbSnSe_bandInversion}, as well as scanning tunneling microscopy and spectroscopy (STM/STS) studies of topological surface states~\cite{Sessi2016_TCI_stepEdges}. Here, high-resolution, bulk-sensitive soft-X-ray ARPES (SX-ARPES; $h\nu = 400-900$~eV)~\cite{Strocov2014_SXARPES} measurements are performed on a high-quality PbSe single crystal. The measurements probe the valence-band dispersions along the X$\Gamma$X, WXW, and K$\Gamma$K directions and the iso-energy surfaces in the $k_z = 0$ plane, resolving the L, $\Sigma$, and $\Delta$ valence-band maxima and their relative energies. First-principles simulations complement the SX-ARPES experiments to map the multivalley electronic structure of PbSe across the \{001\} cross-section of the Brillouin zone. We compare the results of DFT with the PBE and HSE functionals, as well as QP$GW$, to the ARPES data. Of the methods studied here, HSE and QP$GW$ reproduce the measured band dispersions with comparable accuracy, whereas PBE underestimates the band widths, increasingly deviating from experiment for the deeper valence bands. We further compare the results of PBE, HSE, and QP$GW$ based calculations of the Pisarenko relation of p-type PbSe to experimental data~\cite{Wang2011_PbSe_thermoelectric,Chasapis2015_PbSe_twoband}. We find that PBE-based simulations are overall in poor agreement with transport experiments. In the heavily doped regime, where the behavior depends primarily on the band structure, HSE and QP$GW$ are in good agreement with each other and with experiment. In the lightly doped regime, where the band gap is dominates the thermoelectric behavior, only HSE is in good agreement with experiment, whereas PBE underestimates the band gap and QP$GW$ overestimates it.

\section{\label{sec:level2} Methods}

\subsection*{Sample Growth}
A single crystal of PbSe, shown in the inset of Figure~\ref{fig:xrd}(a), was grown by the original method of self-selecting vapor growth (SSVG), successfully employed to grow single crystals of various II-VI and IV-VI families of semiconductor compounds and their substitutional solid solutions, such as Pb$_{1-x}$Sn$_x$Se~\cite{Dziawa2012_TCI_PbSnSe,Sessi2016_TCI_stepEdges}.
Semiconductor materials grown by the SSVG method are of excellent crystal structure as well as good chemical homogeneity and controlled stoichiometry.
In the SSVG method, the crystals grow in a near-equilibrium regime, with the temperature profile of the technological furnace set to ensure growth of the crystal in exclusive contact with its own source material.
In PbSe crystals, controlling slight deviations from stoichiometry between the cation and anion sublattices (i.e., the concentration of electrically active vacancies) permits obtaining the required n- or p-type conductivity~\cite{Szczerbakow2005_SSVG}. The PbSe single crystal studied in this work was grown in a quartz ampoule at a temperature of 817~$^\circ$C, i.e., 261~$^\circ$C below the melting point of PbSe. The growth took 5 weeks and resulted in \{001\}-faceted single-crystalline blocks. The starting material was polycrystalline PbSe synthesized in our technological laboratory in a separate process using 5N-purity metal and nonmetal elements. To remove excess selenium and achieve optimal stoichiometry, the material was additionally annealed at 550~$^\circ$C for 1 hour.

\subsection*{Structural Characterization}
\begin{figure}[h] \centering \includegraphics[width=\linewidth]{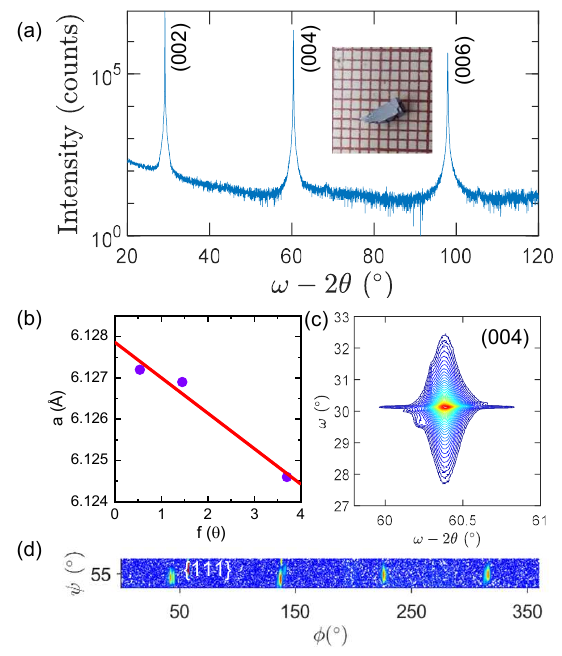} 
\caption{ (a) $\omega$--$2\theta$ XRD pattern of the PbSe sample showing a (001)-oriented single crystalline phase.
The inset shows the size of the single crystal used for ARPES measurements placed on millimeter paper.
(b) Nelson--Riley plot used to determine the lattice constant.
(c) Symmetric (004) reflection map illustrating the crystal perfection of the Bragg peak along the two orthogonal directions, $\omega$ and $\omega$--$2\theta$.
(d) Azimuthal diffraction scan of the asymmetric $\{111\}$ reflections demonstrating the four-fold symmetry of the crystal.}
\label{fig:xrd}
\end{figure}

The high structural quality and well-defined orientation of the crystals are essential for reliable ARPES measurements.
Therefore, the structural quality of the PbSe single crystals was investigated using high-resolution X-ray diffraction (XRD). XRD measurements were performed using a Panalytical X'Pert Pro MRD diffractometer equipped with a Cu~K$\alpha_{1}$ radiation source ($\lambda = 1.54056$~\AA). The incident beam was collimated using an X-ray parabolic mirror and a hybrid two-bounce (220) Ge monochromator. Diffracted X-rays were detected using a PIXcel detector with Soller slits to reduce axial divergence. Measurements were performed at room temperature in scanning mode of the detector with an angular step size of $0.01$--$0.05^{\circ}$. Symmetric $\omega$--$2\theta$ scans, reciprocal space maps, and azimuthal ($\phi$) scans were carried out to analyze the crystal structure, lattice constant, and crystallographic orientation of the PbSe single crystal.
Figure~\ref{fig:xrd}(a) shows the $\omega$--$2\theta$ diffraction pattern, where only the (00$l$) reflections of PbSe are observed, indicating a single-phase crystal with a preferred orientation along the [001] direction.
No additional peaks associated with secondary phases or impurities are detected, confirming the high phase purity of the sample.
The inset of Figure~\ref{fig:xrd}(a) displays the several-millimeter-sized single crystal used for the ARPES measurements.
The lattice constant of the crystal was determined using the Nelson-Riley extrapolation method~\cite{Nelson_1945} with function $\frac{1}{2}\left(\frac{\cos^2\theta}{\sin\theta} + \frac{\cos^2\theta}{\theta}\right)$, as shown in Figure~\ref{fig:xrd}(b).
The extrapolated value of the lattice constant $a = 6.1279 \pm 0.0004$~\AA~is consistent with the reported lattice parameter of PbSe~\cite{Noda1983}.
Further insight into the crystalline quality was obtained from the symmetric reciprocal space map around the (004) reflection shown in Figure~\ref{fig:xrd}(c).
A well-defined Bragg peak is observed, without asymmetry along the $\omega$ and $\omega$--$2\theta$ directions, indicating a small degree of mosaicity (FWHM$_{\omega} = 280''$) and absence of microstrain (FWHM$_{\omega\text{--}2\theta} = 190''$) within the crystal.
The relatively narrow peak distribution further confirms the high crystalline quality of the sample.
Finally, the azimuthal ($\phi$) scan of the asymmetric $\{111\}$ reflections, presented in Figure~\ref{fig:xrd}(d), exhibits a clear four-fold rotational symmetry.
This observation is consistent with the cubic rock-salt crystal structure of PbSe and confirms the well-defined crystallographic orientation of the single crystal.
The excellent structural quality and orientation of the crystal make it suitable for subsequent ARPES measurements.

\subsection*{ARPES}

Synchrotron-radiation ARPES experiments were carried out at the SX-ARPES end station from the ADRESS beamline~\cite{Strocov2014_SXARPES} of the Swiss Light Source installed at beamline P04~\cite{Viefhaus2013_P04} of the PETRA III storage ring at DESY (Hamburg, Germany).
The spectra were recorded over a photon energy ($h\nu$) range of $400$--$900$~eV using circularly polarized light and a hemispherical electron analyzer (PHOIBOS-225, SPECS GmbH).
The beamline and analyzer's combined energy resolution varied from $\sim 100$ to $200$~meV across the investigated $h\nu$ range.
The sample was cleaved in situ and measured at a temperature of 14~K under ultrahigh-vacuum conditions better than $2\times10^{-10}$~mbar.
Certain inhomogeneity of the cleave required adjustment of the beam, focused to about 20~$\mu$m$^{2}$, on the sample surface.
Conversion of the photoelectron kinetic energies and emission angles into crystal momentum values included a correction for the photon momentum~\cite{Strocov2014_SXARPES}.
The coherent (\textbf{k}-resolved) spectral component, superimposed on the Debye-Waller background~\cite{Braun2013_DebyeWaller}, was extracted by subtracting appropriately scaled angle-integrated spectral intensity from the raw ARPES data.
All experimental spectra were aligned in binding energy ($E_b$) by referencing them to the sharp upper edge of the incoherent spectral intensity, which, as discussed in Section~\ref{sec:arpes_comparison}, pinpoints the $\Sigma$~point in the \textbf{k}-integrated density of states (DOS).

\subsection*{Thermoelectric Simulations}\label{sec:te_formalism}

The thermoelectric conversion efficiency of a material is quantified by the dimensionless figure of merit 
\begin{equation}
zT = S^{2}\sigma T/(\kappa_{e} + \kappa_{L})
\end{equation}
where $S$ is the Seebeck coefficient, which measures the voltage generated per unit temperature difference under open-circuit conditions; $\sigma$ is the electrical conductivity; $S^{2}\sigma$ is the power factor, which measures the electrical power a material can generate for a given temperature difference; and $\kappa_{e}$ and $\kappa_{L}$ are the electronic and lattice contributions to the thermal conductivity~\cite{Wang2011_PbSe_thermoelectric,Liu2025_PbSe_TE_review}.

The thermoelectric response of PbSe is treated here within the Boltzmann transport framework, which links the computed band structures, both occupied and unoccupied, to measured transport coefficients.
In this framework, an electron in band $n$ with wave vector $\mathbf{k}$ travels at the band group velocity $\mathbf{v}_{n\mathbf{k}} = \hbar^{-1}\,\partial\varepsilon_{n\mathbf{k}}/\partial\mathbf{k}$ until it is scattered, on average after a relaxation time $\tau$. Within the constant relaxation-time approximation (CRTA), $\tau$ is considered as an energy-independent constant. The transport distribution function sums over the conduction channels available at a given energy, based on the band structure~\cite{Madsen2006_BoltzTraP,Pizzi2014_BoltzWann}:
\begin{equation}
\Xi_{ij}(\varepsilon) = \frac{\tau}{V}\sum_{n\mathbf{k}} v^{i}_{n\mathbf{k}}\,v^{j}_{n\mathbf{k}}\;\delta(\varepsilon - \varepsilon_{n\mathbf{k}}),
\label{eq:tdf}
\end{equation}
where $V$ is the cell volume and $i,j$ are Cartesian directions ($\Xi_{ij} = \Xi\,\delta_{ij}$ in cubic PbSe). The Seebeck coefficient is evaluated based on energy integrals over the transport distribution function:
\begin{equation}
S(\mu, T) = -\frac{1}{eT}\,
\frac{\displaystyle\int d\varepsilon \left(-\frac{\partial f}{\partial \varepsilon}\right) \Xi(\varepsilon)\,(\varepsilon-\mu)}
{\displaystyle\int d\varepsilon \left(-\frac{\partial f}{\partial \varepsilon}\right) \Xi(\varepsilon)},
\label{eq:seebeck}
\end{equation}
where $f(\varepsilon;\mu,T)$ is the Fermi-Dirac distribution, $\mu$ is the chemical potential, and $e$ is the elementary charge. The derivative $-\partial f/\partial\varepsilon$ confines the integrals to an energy window of a few $k_{B}T$ around $\mu$. $S$ vanishes when the states above and below $\mu$ conduct equally, is negative when transport is dominated by states above $\mu$ (electrons), and is positive when it is dominated by states below $\mu$ (holes).
Because $\tau$ appears as a multiplicative factor in $\Xi(\varepsilon)$ in both the numerator and the denominator of Eq.~(\ref{eq:seebeck}), it cancels out in $S$. This means that within the CRTA the Seebeck coefficient is determined solely by the band structure.
In contrast, the conductivity:
\begin{equation}
\sigma = e^{2} \int d\varepsilon \left(-\frac{\partial f}{\partial \varepsilon}\right) \Xi(\varepsilon),
\label{eq:sigma}
\end{equation}
 and the resulting power factor, $S^{2}\sigma$, depend on $\tau$. Evaluating them would thus require an estimate for $\tau$. Therefore, we focus here only on the Seebeck coefficient as a metric for the implications of the accuracy of the computed band structure on the predicted thermoelectric performance.

The Pisarenko relation $S(p_{\mathrm{H}})$ and the temperature-dependent Seebeck coefficient $S(T)$, follow from solving the charge-neutrality condition $p(\mu,T) - n(\mu,T) = p_{\mathrm{H}}$ for the chemical potential $\mu^{*}$. The hole and electron densities $p$ and $n$  are computed from the electronic structure produced by each method (PBE, HSE, and QP$GW$)~\cite{Madsen2006_BoltzTraP}. For $S(p_{\mathrm{H}})$, the temperature $T$ is fixed and the net doping density $p_{\mathrm{H}}$ is varied. For $S(T)$, $p_{\mathrm{H}}$ is fixed at the measured room-temperature Hall density of the pristine sample and the temperature is varied.
The Seebeck coefficient then follows from Eq.~(\ref{eq:seebeck}) as $S = S\big(\mu^{*}, T\big)$ (see further discussion in Section~V of the Supplemental Material (SM)~\cite{SuppMat}).

\subsection*{Computational Details}

All calculations were performed using the Vienna Ab initio Simulation Package (VASP)~\cite{1993_Kresse_AbInitioOrigional_PRB,1996_Kresse_Furthmuller_PRB} with the projector augmented wave (PAW) method~\cite{1994_Blochl_ProjectorAugmentedWaveMethod_PRB,1999_Kresse_UltrasoftPseudopotentialsAumented_PRB}.
The plane-wave basis set kinetic energy cutoff was 400~eV and the energy convergence criterion was $10^{-8}$~eV.
The Brillouin zone was sampled using a $\Gamma$-centered $11\times 11\times 11$ $k$-point mesh.
Gaussian smearing with a width of $\sigma = 0.02$~eV was used.
The PbSe primitive unit cell in the rock-salt structure was used, with a lattice parameter of $a = 6.1213$~\AA, determined by room-temperature ($T = 298$~K) X-ray diffraction~\cite{Noda1987_PbSe_XRD}. This value is within $0.1\%$ of the value determined in Section~\ref{sec:level2} from our XRD data ($a = 6.1279$~\AA).

DFT calculations were performed using the generalized gradient approximation (GGA) of Perdew, Burke, and Ernzerhof (PBE)~\cite{1996_Perdew_GeneralizedGradientApproximation_PRL,PBE_erratum} and the Heyd-Scuseria-Ernzerhof (HSE) hybrid functional~\cite{2003_Heyd_HSE_JChemPhys,2006_Krukau_HSE_JChemPhys}. Quasiparticle self-consistent $GW$ (QP$GW$)~\cite{vanSchilfgaarde2006_QSGW,Shishkin2007_QPGW} calculations were started from PBE wavefunctions. Forty self-consistent iterations were performed to achieve a charge density residual below $10^{-7}$~e/\AA$^3$, using 88 bands and 128 imaginary frequency grid points.
The standard PAW pseudopotentials (\texttt{Pb\_d}, \texttt{Se}) were used for PBE and HSE calculations and GW-specific pseudopotentials (\texttt{Pb\_d\_GW}, \texttt{Se\_GW}) were used for QP$GW$ calculations.
Spin-orbit coupling (SOC) was included in all calculations via the non-collinear formalism implemented in VASP~\cite{Hobbs2000_noncollinear}.

Band structures were obtained via Wannier interpolation using the \texttt{Wannier90} code~\cite{2020_Pizzi_Wannier90_JPCM}.
For PBE and HSE, 26 spinor Wannier functions were constructed directly from the lowest 26 bands, with random projections as initial guesses and 1000 minimization iterations.
For QP$GW$, disentanglement~\cite{Souza2001_disentanglement} was employed to extract 26 Wannier functions from the lowest 36 bands, with a frozen energy window extending to 1.0~eV above the conduction band minimum, so that the occupied manifold and the conduction bands relevant to transport are reproduced exactly.
The interpolated band structures were evaluated along the high-symmetry path $\Gamma$--X--W--L--$\Gamma$--K with 100 points per segment.
Constant-energy contours were obtained by evaluating the Wannier-interpolated band structure on a dense $201\times 201\times 201$ $k$-point grid and then taking the $k_z = 0$ slice, i.e.\ the (001) cross-section of the Brillouin zone.
Thermoelectric transport quantities were computed from these PBE, HSE, and QP$GW$ Wannier Hamiltonians with the \texttt{BoltzWann} module~\cite{Pizzi2014_BoltzWann}, within the CRTA of the Boltzmann transport equation, on a $60\times 60\times 60$ interpolation mesh and with a 50~meV Gaussian broadening for the corresponding density of states. 
The density-of-states effective masses reported in Section~\ref{sec:transport} were obtained by fitting the parabolic-band form $N(\varepsilon) \propto (m_d^{*})^{3/2}\,(\varepsilon_{\mathrm{VBM}} - \varepsilon)^{1/2}$ to each method's computed density of states within 0.15~eV below the valence-band maximum $\varepsilon_{\mathrm{VBM}}$.

\section{\label{sec:level3} Results and discussion}

\subsection{Simulated Band Structure}

\begin{figure}[!htbp]
\centering
\includegraphics[width=0.95\linewidth]{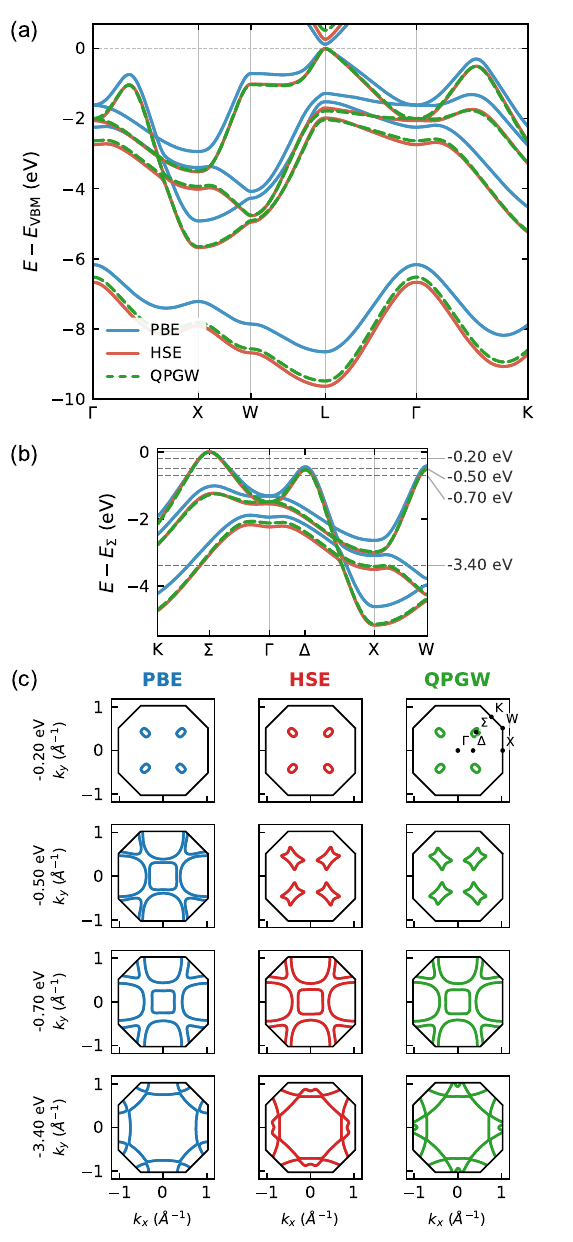}
\caption{Computed valence band structure of PbSe.
(a)~Band structure along the high-symmetry path $\Gamma$--X--W--L--$\Gamma$--K obtained with PBE (blue, solid), HSE (red, solid), and QP$GW$ (green, dashed).
Energies are referenced to the VBM at the L-point.
(b)~Magnified view along K--$\Gamma$--X--W. Energies are referenced to the $\Sigma$ band maximum along $\Gamma$--K, which is the local VBM in the (001) plane. The horizontal dashed lines at $-0.2$, $-0.5$, $-0.7$, and $-3.4$~eV indicate the four iso-energy cuts displayed in (c).
(c)~Constant-energy contours of the bulk band structure with $k_z = 0$. The rows correspond to energies of $-0.2$, $-0.5$, $-0.7$, and $-3.4$~eV below the $\Sigma$ band maximum along $\Gamma$--K, indicated in panel (b), and the columns correspond to the PBE, HSE, and QP$GW$ methods. The octagonal boundary denotes the (001) surface Brillouin zone. 
}
\label{fig:band_and_iso}
\end{figure}

Figure~\ref{fig:band_and_iso} shows the computed band structure of PbSe. The band structures obtained using PBE, HSE, and QP$GW$ are qualitatively similar. The valence band structures obtained using HSE and QP$GW$ are very close to each other. In comparison, the PBE band structure appears compressed with the band width underestimated and the bands shifted closer to the Fermi level. The differences between the methods become more apparent farther from the Fermi level.

All three methods produce a direct band gap at the L~point, as established for the rock-salt phase~\cite{Dalven1973_PbSeBandGap,Wei1997_PbSeBS}.
PBE yields an inverted band gap of $E_g = 0.12$~eV, underestimating the experimental room-temperature gap of $\approx 0.27$~eV~\cite{Dalven1973_PbSeBandGap}. This is a consequence of the self-interaction error (SIE) affecting semi-local exchange-correlation functionals, known to manifest for the narrow-gap Pb chalcogenides~\cite{Hummer2007_PbChalcogenides_GW}. HSE yields a gap of $0.26$~eV, reproducing the experimental gap within $\sim 10$~meV thanks to the fraction of  Fock exchange, which mitigates the SIE~\cite{2003_Heyd_HSE_JChemPhys,Paier2006_HSE_benchmark,Paier2006_HSE_erratum}. QP$GW$ produces a gap of $0.52$~eV, overestimating the gap by about a factor of two. This may be attributed to the neglect of the electron-hole vertex in the screened Coulomb interaction, $W$, which underestimates the screening and widens the quasiparticle gap, an error that is most significant for narrow-gap materials~\cite{Shishkin2007_vertex,Grumet2018_QPGW_overestimate}. It has been shown that non-self-consistent $G_0W_0$ and partially self-consistent $GW$ schemes can produce band gap values closer to experiment thanks to a fortuitous error cancellation between the over-screening of the semi-local DFT starting point and the under-screening of $GW$~\cite{Grumet2018_QPGW_overestimate}. Our results are consistent with Ref.~\cite{Hummer2007_PbChalcogenides_GW}.

The valence band maximum (VBM) is located at the L-point. The topmost valence band at L is doubly degenerate (Kramers pair) and, in HSE and QP$GW$, derives primarily from Se~$4p$ states hybridized with Pb~$6s$~\cite{Wei1997_PbSeBS,Svane2010_QPGW_PbChalcogenides} (orbital-resolved band structures for all three methods are provided in Section~IV of the SM~\cite{SuppMat}). This even-parity $L_6^+$ state forms the VBM, and the odd-parity $L_6^-$ state of mixed Se~$4s$--Pb~$6p$ character forms the conduction-band minimum (CBM). PBE inverts this ordering, a topological discrepancy consistent with the inverted gaps reported for semi-local functionals~\cite{Hummer2007_PbChalcogenides_GW,Svane2010_QPGW_PbChalcogenides}.
The second highest extremum is the $\Sigma$ point along $\Gamma-K$, whose position below the VBM is $0.30$~eV, $0.50$~eV, and $0.51$~eV based on PBE, HSE, and QP$GW$, respectively. The $\Sigma$ valley carries a valley degeneracy of $N_v = 12$, compared with $N_v = 4$ at L.
For heavily hole doped PbSe at elevated temperatures, the 12 $\Sigma$ valleys contribute to the transport in addition to the 4 L valleys, enhancing the Seebeck coefficient~\cite{Zhao2017_ARPES_PbSe_convergence, Pei2011_PbSeConvergence,Wang2011_PbSe_thermoelectric,Chasapis2015_PbSe_twoband,Zhu2022_multiband}.
The $\sim 200$~meV spread of the $L-\Sigma$ offset across methods contributes to differences in the predicted thermoelectric properties, as discussed in Section~\ref{sec:transport} and in Section~V of the SM~\cite{SuppMat}.
The topmost band also has a local maximum at the $\Delta$ point along $\Gamma-X$ ($\approx 0.34 \times 2\pi/a$; Figure~\ref{fig:BZ}), which is located $0.74$, $1.04$, and $1.05$~eV below the VBM with PBE, HSE, and QP$GW$, respectively (Figure~S4 of the SM~\cite{SuppMat}). 
At that point, the Se~$4p$-dominated manifold develops a weak Pb~$6s$ hybridization, whose weight increases toward the X point (see the orbital decomposition in Section~IV of the SM~\cite{SuppMat})~\cite{Wei1997_PbSeBS,Hummer2007_PbChalcogenides_GW}.
A well-separated lower valence band of mixed Pb~$6s$--Se~$4s$ bonding character lies $\sim 7$--$9$~eV below the VBM.

\begin{figure*}[tbp]
\centering
\includegraphics[width=0.95\textwidth]{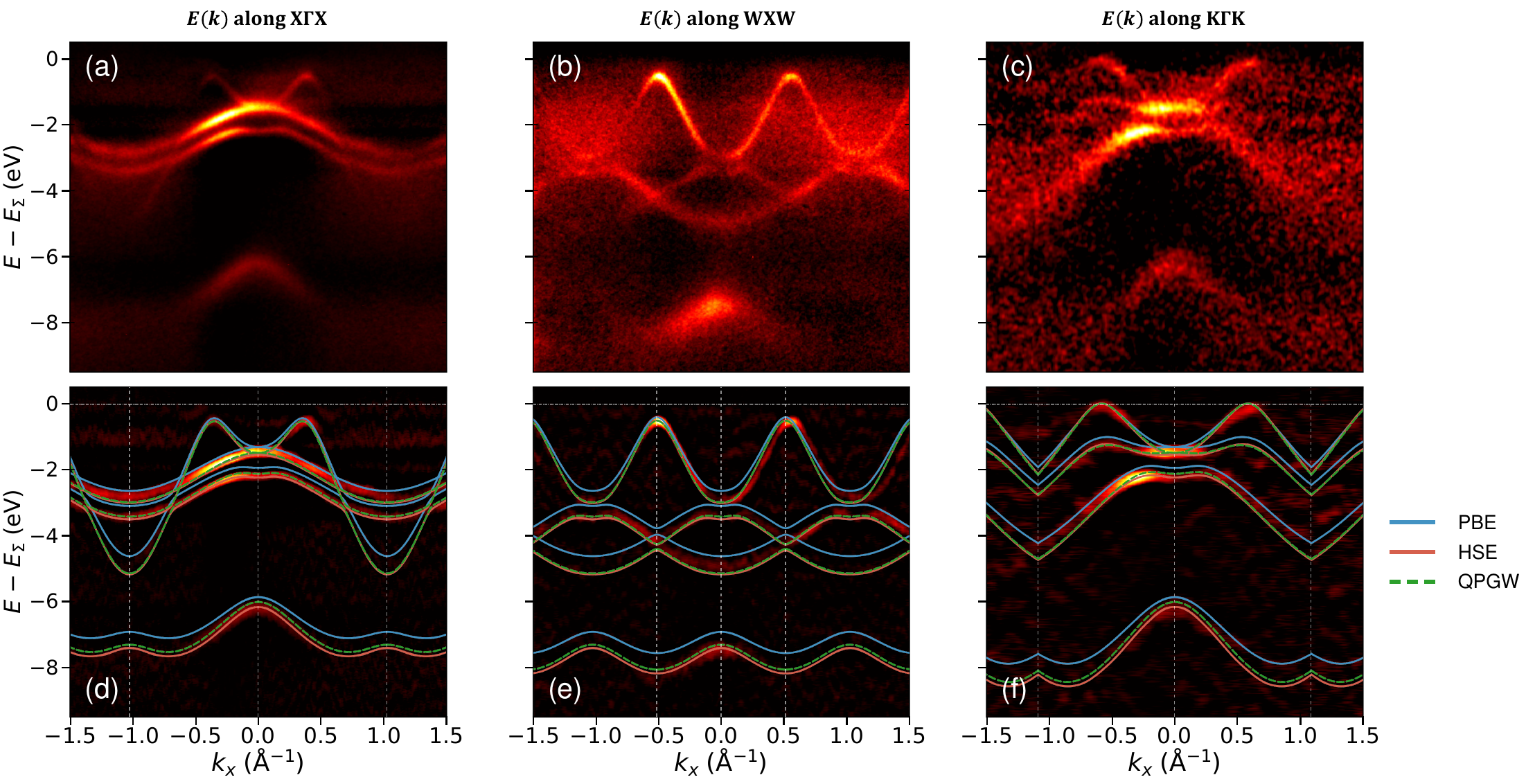}
\caption{ARPES intensity of PbSe, measured along (a) X$\Gamma$X, (b) WXW, and (c) K$\Gamma$K. The clipped negative second derivative of the ARPES intensity with respect to energy, $-\partial^{2}I/\partial E^{2}$, compared to the PBE, HSE, and QP$GW$ band structures along (d) X$\Gamma$X, (e) WXW, and (f) K$\Gamma$K.
All $E_b$ scales are referenced to the $\Sigma$~band maximum. The vertical dashed lines mark the high-symmetry along each line. 
}
\label{fig:band_overlay}
\end{figure*}

Figure~\ref{fig:band_and_iso}(c) shows constant-energy contours on the $k_z = 0$ cross-section at $E - E_\Sigma = -0.2$, $-0.5$, $-0.7$, and $-3.4$~eV, where $E_\Sigma$ denotes the $\Sigma$ band maximum along $\Gamma-K$.
In this plane, the L valleys project onto four equivalent pockets elongated along $\langle 110 \rangle$. 
The differences between methods are most apparent in the $E - E_\Sigma = -0.5$~eV and the $E - E_\Sigma = -3.4$~eV iso-contours.
The $-0.5$~eV iso-energy contour differs between methods because of its position relative to the two maxima of the topmost band. The W-point maximum is located at $E - E_\Sigma = -0.42$~eV with PBE but at $-0.51$ and $-0.52$~eV with HSE and QP$GW$, respectively. The $\Delta$ maximum lies at $-0.44$~eV with PBE versus $-0.53$ and $-0.54$~eV with HSE and QP$GW$, respectively (Figure~\ref{fig:band_and_iso}(b)).
At energies below the W-point maximum, the four projected-L pockets connect through the W points at the corners of the octagonal zone boundary. Below the $\Delta$ maximum, an additional closed pocket forms around $\Gamma$.
The $-0.5$~eV iso-energy contour lies below both maxima with PBE due to the compressed band width, whereas with HSE and QP$GW$ it lies above both maxima. Therefore, the HSE/QP$GW$ contours still show four separate pockets. This has implications for transport, as discussed in Sections~\ref{sec:arpes_comparison} and~\ref{sec:transport} below.
Another significant difference between PBE and HSE/QP$GW$ appears at the $-3.4$~eV depth, and its origin is also evident from Figure~\ref{fig:band_and_iso}(b). The $E - E_\Sigma = -3.4$~eV dashed line skims the nearly-flat second-from-top valence band at X, of mixed Se~$4p$--Pb~$6p$ character, in the HSE and QP$GW$ band structures.
In the PBE band structure, the same band lies about $\sim 0.4$~eV higher in energy, such that the $-3.4$~eV line already lies below the band at X and crosses it obliquely along X--W. At this energy depth, the HSE and QP$GW$ band structures also differ slightly around the X-point.
Apart from the band ordering at L discussed above, the differences between the methods are therefore mostly quantitative and affect the energy scale more than other aspects. For instance, the two anticrossings between the three uppermost valence bands along $\Gamma$--X occur, though at different energies, at the same momenta in all three methods to within 0.02~\AA$^{-1}$.

\subsection{\label{sec:arpes_comparison}Comparison with ARPES }

\begin{figure*}[tbp]
\centering
\includegraphics[width=0.95\textwidth]{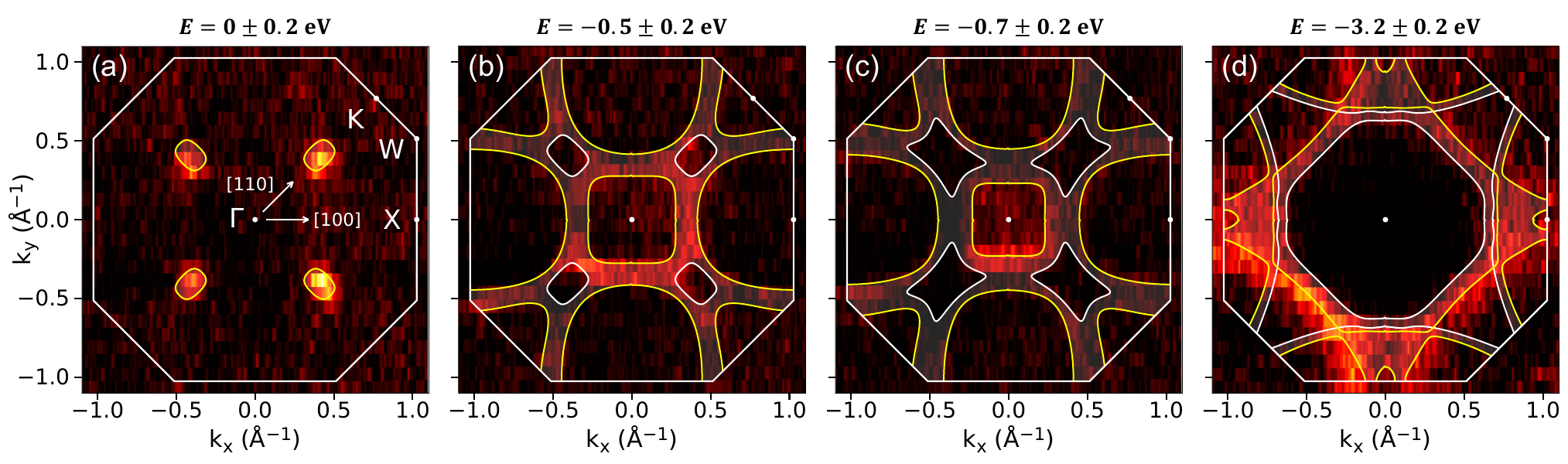}
\caption{Comparison of the QP$GW$ constant-energy contours (white and yellow lines) with ARPES iso-energy maps in the $k_z = 0$ plane at four $E_b$ values, referenced to the local VBM at $\Sigma$: $E - E_\Sigma =$  (a) $0\pm0.2$, (b) $-0.5\pm0.2$, (c) $-0.7\pm0.2$, and (d)$-3.2\pm0.2$~eV.
In each panel, the ARPES intensity is integrated over an $E_b$ window of $\pm 0.2$~eV. The white contour marks the top edge of the integration window and the yellow contour marks its bottom edge (in panel (a) the $+0.2$~eV top edge lies above the in-plane valence-band maximum $E_\Sigma$, so only the bottom-edge contour appears). The region between the contours is shaded. The octagonal boundary marks the (001) surface Brillouin zone.}
\label{fig:iso_overlay}
\end{figure*}

\begin{figure*}[tbp]
\centering
\includegraphics[width=0.95\textwidth]{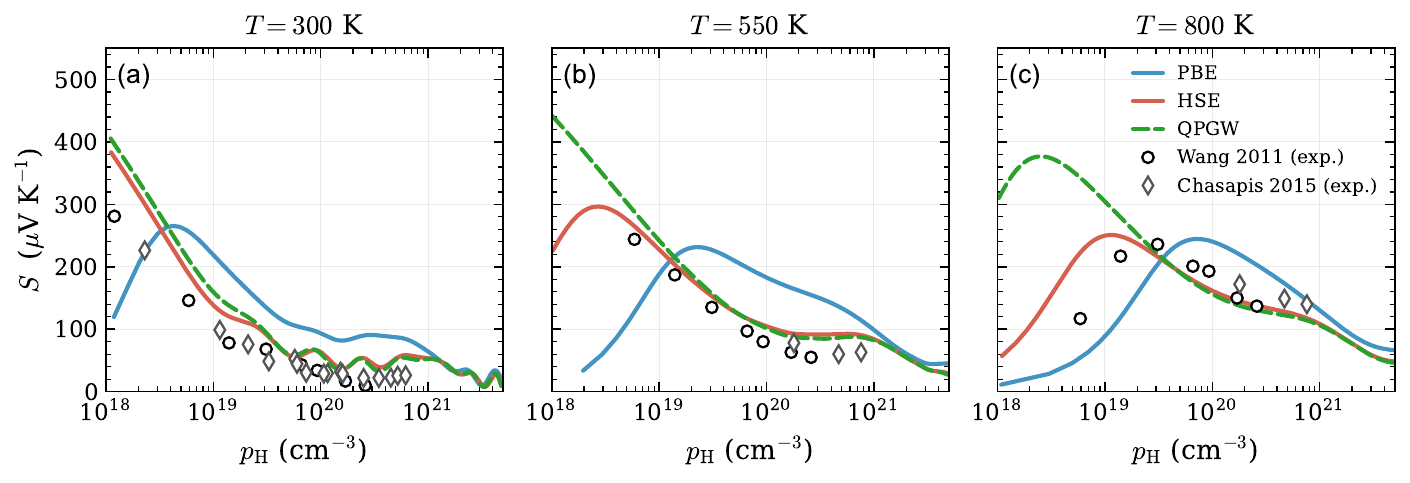}
\caption{Pisarenko relation of p-type PbSe at temperatures of (a) 300~K, (b) 550~K, and (c) 800~K, calculated based on the PBE, HSE, and QP$GW$ Wannier Hamiltonians (CRTA), compared to the Pb$_{1-x}$Na$_x$Se samples of Ref.~\cite{Wang2011_PbSe_thermoelectric} (circles) and to the Pb$_{1-x}$Na$_x$Se ingot and spark-plasma-sintered (SPS) samples of Ref.~\cite{Chasapis2015_PbSe_twoband} (diamonds). For the experimental points, the abscissa $p_{\mathrm{H}}$ is the measured Hall density of each sample (taken at room temperature or below). The computed curves are calculated over the net doping density $p_{\mathrm{H}} = p(\mu,T) - n(\mu,T)$ with no experimental input.
}
\label{fig:pisarenko}
\end{figure*}

Figure~\ref{fig:band_overlay} shows ARPES dispersions measured along the surface-parallel X$\Gamma$X, WXW, and K$\Gamma$K directions (see the Brillouin zone illustration in Figure~\ref{fig:BZ}) as raw intensity images (top row) and overlaid with the three theoretical band structures (bottom row).
To facilitate the comparison between theory and experiment, the ARPES images were denoised by Gaussian smoothing (0.35~eV FWHM) in the energy direction. Subsequently, the dispersions were enhanced by representing the ARPES intensity $I(E,\mathbf{k})$ by the negative second derivative $-\partial^{2} I / \partial E^{2}$, with negative values clipped to zero.
The spectra show a high signal-to-background ratio, although the statistics of the K$\Gamma$K image is somewhat limited by a sporadic beam dump during the acquisition.
The binding-energy scale of all three momentum cuts is referenced to the local maximum at the $\Sigma$~point rather than to the Fermi level. The reason for this choice is that the Fermi level in pristine PbSe is set by native defects (Pb/Se vacancies)~\cite{Ravich1970_LeadChalcogenides,Qin2024_Science_gridplain}. At the carrier densities encountered here ($\sim (1$--$5)\times 10^{18}$~cm$^{-3}$), the Fermi level lies about $0.03$--$0.1$~eV below the L~VBM, drifting with carrier concentration and temperature. In contrast, the band maximum at $\Sigma$, is a deeper feature that is revealed clearly by ARPES~\cite{Zhao2017_ARPES_PbSe_convergence} and is visible in Figure~\ref{fig:band_overlay} as the sharp upper edge of the incoherent (\textbf{k}-integrated) intensity. Therefore, the band maximum at $\Sigma$ is a more well-defined reference point than the Fermi level.

The dispersions shown here, measured as $E(k_x)$, are genuinely of bulk origin. In the SM~\cite{SuppMat} (Figure~S1), the $E(k_z)$ dispersion along $\Gamma$--X (obtained by scanning the photon energy) is compared with the $E(k_x)$ dispersion along the same direction. The two cuts show identical valence-band features at the same $E_b$, demonstrating the bulk sensitivity of soft-x-ray ARPES at the photon energies used here ($400$--$900$~eV)~\cite{Strocov2014_SXARPES}.

The HSE and QP$GW$ band structures agree with the ARPES bands to within 0.1--0.2~eV over the entire valence band, as detailed below.
The maxima at the $\Delta$-point and W-point, both measured at $E - E_\Sigma \approx -0.5$~eV, lie on top of the HSE and QP$GW$ bands, whose maxima at $-0.51$ and $-0.53$~eV (HSE) and at $-0.52$ and $-0.54$~eV (QP$GW$) agree with the measured value within the experimental energy resolution, whereas the PBE bands are shallower in energy by $\sim 0.15$~eV. The deviation of PBE from experiment grows with the distance from the band-edge region and reaches $\sim 0.3$--$0.4$~eV at the $\Gamma$ and X extrema of the topmost band.
The same trends are observed for the intermediate binding-energy manifold. The HSE and QP$GW$ bands remain within $\sim 0.1$~eV of each other and closely trace the ARPES bands, whereas the corresponding PBE bands are shifted toward shallower binding energies by $0.3$--$0.6$~eV.
For the well-separated lower valence band of mixed Pb~$6s$--Se~$4s$ character~\cite{Wei1997_PbSeBS}, the QP$GW$ and HSE  bands bracket the ARPES bands from above and below, respectively, to within $\sim 0.1$--$0.2$~eV, whereas the PBE bands are too shallow by $\sim 0.4$--$0.5$~eV. 
Taken together, these deviations show that the PBE band structure is compressed toward the band edge relative to ARPES, as noted in the discussion of Figure~\ref{fig:band_and_iso}(a).

It is worth noting that all calculations refer to a static lattice at the room-temperature lattice constant, whereas the ARPES measurements were performed at 14~K. The room-temperature value was used because no structural characterization data of our sample are available at the ARPES temperature, and because correcting it with the thermal expansion measured on other samples~\cite{Knight2022_PbSe_thermoelastic} would introduce additional experimental priors and extrapolation errors into the calculations and compromise their first-principles nature. Interestingly, despite this temperature difference, the agreement of all band dispersions along the three cuts (and of the iso-energy contours discussed below) indicates that the deformation-potential shifts and the electron-phonon renormalization of the valence bands in the $k_z = 0$ plane are relatively weak, in contrast to the more temperature-sensitive L--$\Sigma$ separation~\cite{Zhao2017_ARPES_PbSe_convergence,Chasapis2015_PbSe_twoband,Querales2019_PbTe_Tbands} and band gap~\cite{Dalven1973_PbSeBandGap,Gibbs2013_PbX_Tdependent_gap}.

In Figure~\ref{fig:iso_overlay}, the  iso-energy contours calculated with QP$GW$ are compared to ARPES. The ARPES intensity was integrated over a window of $\pm 0.2$~eV to account for the experimental energy resolution of $\approx 120$~meV FWHM.
At $E - E_\Sigma = 0$~eV (Figure~\ref{fig:iso_overlay}(a)), the ARPES intensity is concentrated in four bright spots centered at the $\Sigma$ points on the $\Gamma$--K lines, $\approx (\pm 0.41, \pm 0.41)\times 2\pi/a$. They correspond to the projected L valleys, which are centered at $(\pm 0.5, \pm 0.5, \pm 0.5)\times 2\pi/a$ (Figure~\ref{fig:BZ}) and whose expansion with binding energy intersects the $k_z = 0$ plane around the $\Sigma$ points.
Each QP$GW$ contour encloses one of these spots. The contours are shaped like an elongated ellipse whose major axis lies along $\langle 110 \rangle$, reflecting the prolate anisotropy of the L valley, whose longitudinal axis is oriented along $\langle 111 \rangle$~\cite{Mitchell1966_PbSe_kp,Dalven1973_PbSeBandGap,Svane2010_QPGW_PbChalcogenides}.
A residual asymmetry in the ARPES intensity among the four symmetry-equivalent pockets, most visible in the lower-right pocket, may be attributable to photoemission matrix-element effects, which depend on the direction of the photon momentum relative to the pocket orientation~\cite{Moser2017_ARPES_matrixElements}. 

At $E - E_\Sigma = -0.5 \pm 0.2$~eV (Figure~\ref{fig:iso_overlay}(b)), the bright regions expand from the projected L-pocket centers and merge into continuous bright intensity ridges that form a square along the four $\langle 100 \rangle$ directions.
This evolution gives rise to a closed pocket at $\Gamma$ and to eight ``radiating'' segments that extend from the $\Sigma$ pockets toward the W points.
 These continuous momentum-space bridges between the valleys become accessible to holes in heavily doped p-PbSe. The states on these bridges then open inter-valley scattering channels, which erodes the mobility advantage of the multivalley band structure~\cite{Park2021_convergence}.
The QP$GW$ window spans  this transition, because it straddles the W- and $\Delta$-point maxima of the topmost band at $\sim -0.5$~eV (Figure~\ref{fig:band_and_iso}(b)).
At $E - E_\Sigma = -0.7 \pm 0.2$~eV (Figure~\ref{fig:iso_overlay}(c)), the four $\Sigma$ pockets expand, the $\Gamma$-centered closed pocket shrinks, and the bright ridges connecting the projected L valleys broaden.
The QP$GW$ contours track each of these changes. As seen in the band dispersion in Figure~\ref{fig:band_and_iso}(b), these deeper cuts intersect the topmost band closer to $\Gamma$, and further below the $\Sigma$ maxima, widening the connected ridges.

At $E_b$, $E - E_\Sigma = -3.2 \pm 0.2$~eV (Figure~\ref{fig:iso_overlay}(d)), the intensity redistributes into an approximately octagonal bright ring that follows the BZ boundary between the X and W points, while the central $\Gamma$ region remains dark.
The QP$GW$ contours bracket this pattern. The valence bands at this depth disperse further downward toward the zone boundary (Figure~\ref{fig:band_and_iso}(b)), such that the shallow edge of the window intersects them near the anti-crossing at about two-thirds along $\Gamma$--X, while its deep edge captures the nearly-flat band at X. This match indicates that the more dispersive Se~$4p$--Pb~$6p$ valence band at this depth~\cite{Wei1997_PbSeBS,Svane2010_QPGW_PbChalcogenides} also has the same topology and reciprocal-space extent in QP$GW$ as in experiment.
Taken together, the four panels show that QP$GW$ reproduces the full two-dimensional valence-band structure in quantitative agreement with the experimental momentum-space intensity over an energy window of $\sim 4$~eV.

Comparisons of the PBE and HSE iso-energy contours with the same ARPES maps are provided in Figure~S3 of the SM~\cite{SuppMat}. The PBE and HSE contours are clearly distinguishable from the QP$GW$ contours at selected energies, while remaining qualitatively similar at the others. For instance, at $-0.7$~eV the PBE contour at the upper edge of the integration window already reaches the zone boundary, and at $-3.2$~eV the PBE contour around $\Gamma$ is too large, whereas HSE does not reproduce the X pocket. A more detailed discussion is given in the SM~\cite{SuppMat}. 
Out-of-plane iso-energy maps in the vertical $\Gamma$XW plane, measured at the same four binding energies and shown in Figure~S2 of the SM~\cite{SuppMat}, exhibit the same qualitative evolution and similar agreement with the QP$GW$ contours. This is consistent with the bulk sensitivity of the soft-X-ray ARPES experiment.

\subsection{\label{sec:transport}Thermoelectric response}

PbSe is a leading medium-temperature thermoelectric material whose transport response is governed by the band-structure features discussed above~\cite{Pei2011_PbSeConvergence,Wang2011_PbSe_thermoelectric,Parker2010_PbSe_thermoelectric}. We compare the results of PBE, HSE, and QP$GW$ to transport measurements for PbSe. The Pisarenko relation is the change in the Seebeck coefficient as a function of hole concentration $S(p_{\mathrm{H}})$ at a fixed temperature. Figure~\ref{fig:pisarenko} compares the Pisarenko relation, calculated using PBE, HSE, and QP$GW$, with the measured Seebeck data of undoped and Na-doped PbSe~\cite{Wang2011_PbSe_thermoelectric,Chasapis2015_PbSe_twoband}.
These experiments were chosen because they span the full range of carrier concentration, from the lightly doped regime to the heavily doped, degenerate regime. Because the Seebeck coefficient is independent of the relaxation time within the CRTA, as described in Section~\ref{sec:level2}, the computed curves contain no experimental input and no fitted parameters.

In the degenerate regime ($p_{\mathrm{H}} \gtrsim 10^{19}$~cm$^{-3}$, where bipolar physics is frozen out), the differences between methods are governed by the L--$\Sigma$ offset and the L-valley effective mass. With PBE, the $\Sigma$ maximum lies only $0.30$~eV below the L~VBM, compared to $0.50$--$0.51$~eV with HSE/QP$GW$ (see Figure~\ref{fig:band_and_iso}). As a result, with PBE, the $N_v = 12$ $\Sigma$ valley contributes to the thermally weighted density of states already at moderate doping. In addition, with PBE the L~band is flatter than with HSE/QP$GW$, leading to a larger density-of-states effective mass of $m_d^{*} = 0.44\,m_e$, compared to $0.32\,m_e$ with HSE/QP$GW$ (see Figure~\ref{fig:band_and_iso}). Together, the underestimated L--$\Sigma$ offset and the overestimated L-valley effective mass, make the PBE Seebeck coefficient exceed the HSE/QP$GW$ values by a factor of $\sim 1.7$ at every temperature.

In the degenerate doping regime, the HSE and QP$GW$ curves are within $10\%$ of each other. Both agree reasonably well with the measured data, correctly tracking the dependence of the Seebeck coefficient on the hole concentration across different temperatures. The Seebeck coefficients computed with HSE and QP$GW$ are somewhat overestimated at 300~K and somewhat underestimated at 800~K.
The overestimation at 300~K can be partially attributed to the energy dependence of the relaxation time, $\tau(E)$, which is neglected here. It has been shown that the relaxation time can be approximately described by the acoustic deformation-potential form: $\tau(E) \propto E^{-1/2}$, which reduces $S$ to two-thirds of the constant-$\tau$ value for a degenerate parabolic band~\cite{Bardeen1950_DeformationPotential,Ravich1970_LeadChalcogenides,Ravich1971_ScatteringPSSB,Wang2011_PbSe_thermoelectric}. The underestimation at 800~K may be attributed to neglecting the thermal evolution of the band structure. It has been shown that with increasing temperature, the band-edge mass grows and the $\Sigma$ band shifts toward the L~VBM, both of which raise the measured Seebeck coefficient~\cite{Wang2011_PbSe_thermoelectric,Zhao2017_ARPES_PbSe_convergence,Chasapis2015_PbSe_twoband}.

In the low doping regime ($p_{\mathrm{H}} \lesssim 10^{19}$~cm$^{-3}$) the behavior of $S(p_{\mathrm{H}})$ is dominated by the band gap, $E_g$. Thermally excited minority electrons compensate the hole thermopower, and $S(p_{\mathrm{H}})$ passes through a bipolar maximum whose position depends exponentially on $E_g/k_BT$. Therefore, at a sufficiently low acceptor density, $S$ decreases as $T$ increases (the temperature dependence of the Seebeck coefficient is discussed in detail in Section~V of the SM~\cite{SuppMat}). The hole concentration at which the maximum of $S(p_{\mathrm{H}})$ is reached increases with temperature. Because PBE yields a significantly underestimated band gap of 0.12~eV, the position of the maximum of $S(p_{\mathrm{H}})$ is overestimated. In contrast, the overestimated gap of 0.52~eV produced by QP$GW$, leads to a considerable underestimation of the position of the maximum. The HSE band gap of 0.26~eV is very close to the experimental value of 0.27~eV. Therefore, the position of the $S(p_{\mathrm{H}})$ maximum obtained with HSE is the closest to experiment. This shows that using an electronic structure method that accurately describes the band structure and the band gap is essential to correctly predicting the carrier concentration dependence of a material's thermoelectric response.

\section{\label{sec:level4} Conclusion}

In summary, the valence band structure of a high-quality PbSe single crystal was probed by bulk-sensitive soft-X-ray ARPES. $E(\mathbf{k})$ dispersions were measured along X$\Gamma$X, WXW, and K$\Gamma$K, and iso-energy maps were collected in the horizontal ($k_z = 0$) and vertical ($k_y = 0$) $\Gamma$XW planes. Comparison to the ARPES data enabled us to conduct a rigorous assessment of the performance of DFT with the semi-local PBE functional and the HSE hybrid functional, as well as the beyond-DFT QP$GW$ method within the framework of many-body perturbation theory.
We find that, owing the self-interaction error, PBE underestimates the band width and produces a ``compressed" band structure, whose deviation from experiment increases with depth below the VBM.  
HSE and QP$GW$ reproduce the ARPES dispersions quantitatively throughout the valence manifold, including the relative positions of the local maxima at the $\Sigma$ and $\Delta$ points. The QP$GW$ iso-energy contours also closely reproduce the measured band topology.

We further investigated the implications of the band structure features for the thermoelectric response of PbSe by comparing the Pisarenko relation, $S(p_{\mathrm{H}})$, computed using PBE HSE, and QP$GW$ to experimental data at different temperatures. In the heavily doped regime, $S(p_{\mathrm{H}})$ is very sensitive to the details of the valence band structure. Therefore, in this regime, the results of HSE and QP$GW$ are in agreement with each other and with experiment, whereas PBE significantly overestimates $S(p)$. In the lightly doped regime, $S(p_{\mathrm{H}})$ is sensitive to the band gap. Therefore, HSE, which produces the closest band gap to experiment, correctly tracks the behavior of $S(p_{\mathrm{H}})$. PBE, which underestimates the band gap, overestimates the concentration at which $S(p_{\mathrm{H}})$ reaches the bipolar maximum, whereas QP$GW$, which overestimates the band gap, underestimates the concentration at which $S(p_{\mathrm{H}})$ reaches a maximum. 

Based on the results presented here, we conclude that in order to correctly predict the carrier concentration dependence of the thermoelectric response, it is vital to use an electronic structure method that accurately reproduces both the band structure and the band gap. This has implications for efforts to computationally discover new thermoelectric materials.

\section{\label{sec:level5} Acknowledgments}
 Research at Carnegie Mellon University and the University of Pittsburgh was funded by the U.S. Department of Energy through grant DE-SC0019274.
This research used resources of the National Energy Research Scientific Computing Center (NERSC), a DOE Office of Science User Facility supported by the Office of Science of the U.S.
Department of Energy under contract no.
DE-AC02-05CH11231. 
 The research of V.V.V, J.K and T.S.  was partially supported by the Foundation for Polish Science project ”MagTop” no.
FENG.02.01-IP.05-0028/23 co-financed by the European Union from the funds of Priority 2 of the European Funds for a Smart Economy Program 2021-2027 (FENG).
V.V.V. also acknowledges Narodowe Centrum Nauki (NCN, National Science Centre, Poland) IMPRESS-U Project No. 2023/05/Y/ST3/00191 and long-term program of support of the Ukrainian research teams at the Polish Academy of Sciences carried out in collaboration with the U.S.
National Academy of Sciences with the financial support of external partners. 

\section*{Data Availability}
The input and output files of the calculations and the SX-ARPES data presented in this work are available through Zenodo at \url{https://zenodo.org/records/22286848} (DOI: 10.5281/zenodo.22286848).

\bibliography{reference}

\end{document}


\title{\Large \textbf{Supplemental Material: Multivalley 3D Electronic Structure of PbSe from Soft-X-Ray ARPES and First-Principles Calculations}}

\author{Zefeng Cai}
\affiliation{Department of Materials Science and Engineering, Carnegie Mellon University, Pittsburgh, PA 15213, USA}
\author{Valentine V. Volobuev }
\affiliation{International Research Centre MagTop, Institute of Physics, Polish Academy of Sciences, 02668 Warsaw, Poland}
\affiliation{National Technical University ``KhPI'', 61002 Kharkiv, Ukraine}
\author{Jędrzej Korczak}
\affiliation{International Research Centre MagTop, Institute of Physics, Polish Academy of Sciences, 02668 Warsaw, Poland}
\affiliation{Institute of Physics, Polish Academy of Sciences, 02668 Warsaw, Poland}
\author{Enrico Della Valle}
\affiliation{Paul Scherrer Institut, Swiss Light Source, CH-5232 Villigen PSI, Switzerland}
\author{Hantian Liu}
\affiliation{Department of Materials Science and Engineering, Carnegie Mellon University, Pittsburgh, PA 15213, USA}
\author{Moritz Hoesch}
\affiliation{Deutsches Elektronen-Synchrotron DESY, 22607 Hamburg, Germany}
\author{Sergey M. Frolov}
\affiliation{Department of Physics and Astronomy, University of Pittsburgh, Pittsburgh, PA, 15260, USA} 
\author{Tomasz Story }
\email{Electronic mail: story@ifpan.edu.pl}
\affiliation{International Research Centre MagTop, Institute of Physics, Polish Academy of Sciences, 02668 Warsaw, Poland}
\affiliation{Institute of Physics, Polish Academy of Sciences, 02668 Warsaw, Poland}
\author{Vladimir N. Strocov}
\thanks{Electronic mail: vladimir.strocov@psi.ch}
\affiliation{Paul Scherrer Institut, Swiss Light Source, CH-5232 Villigen PSI, Switzerland}
\author{Noa Marom}
\email{Electronic mail: nmarom@andrew.cmu.edu}
 \affiliation{Department of Materials Science and Engineering, Carnegie Mellon University, Pittsburgh, PA 15213, USA}
\affiliation{Department of Chemistry, Carnegie Mellon University, Pittsburgh, PA 15213, USA}
\affiliation{Department of Physics, Carnegie Mellon University, Pittsburgh, PA 15213, USA}

\date{\today}

\maketitle

\tableofcontents

\newpage

\section{Out-of-plane ARPES measurements}

\begin{figure}[H]
\centering
\includegraphics[width=0.5\linewidth]{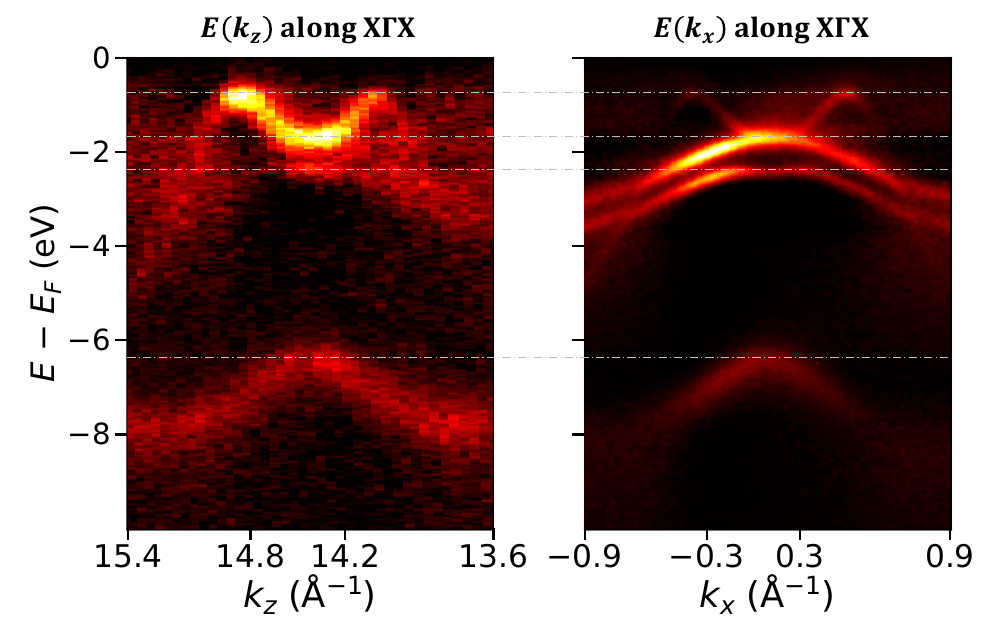}
\caption{ARPES intensity along $\Gamma$--X obtained by scanning $h\nu$ and converting the photoelectron kinetic energy to $k_z$ via the free-electron final-state approximation (left) and by direct imaging of $E(k_x)$ at fixed $h\nu$ (right).
The grey dashed lines mark the binding energies of characteristic valence-band points ($\sim -1$, $-2$, $-3$, and $-6$~eV).
Energies are calibrated relative to $E_F$.}
\label{fig:SI_kzkx}
\end{figure}

\begin{figure}[H]
\centering
\includegraphics[width=0.95\linewidth]{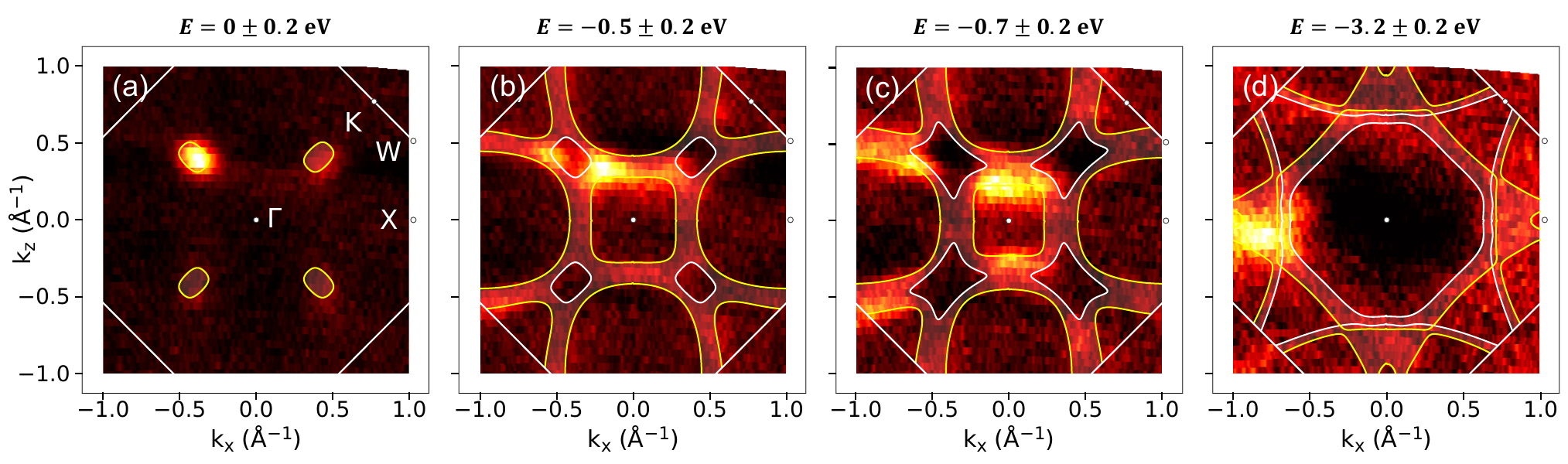}
\caption{Out-of-plane ARPES iso-energy maps in the vertical $\Gamma$XW plane overlaid with the QP$GW$ constant-energy contours (white and yellow lines, marking the top and bottom edges of the integration window as in Figure~5 of the main text), at the same four binding energies as Figure~5 of the main text: $E - E_\Sigma = 0 \pm 0.2$, $-0.5 \pm 0.2$, $-0.7 \pm 0.2$, and $-3.2 \pm 0.2$~eV ((a)--(d)).}
\label{fig:SI_oop_iso}
\end{figure}

Out-of-plane $E(k_z)$ cuts and iso-energy maps in the vertical $\Gamma$XW plane ($k_y = 0$) were recorded under variation of $h\nu$.
Figure~\ref{fig:SI_kzkx} compares the $E(k_z)$ and $E(k_x)$ dispersions along the equivalent $\Gamma$--X directions displayed in ARPES intensity (without the $-\partial^2 I / \partial E^2$ post-processing done for the overlay shown in Figure~4 in the main text).
Because each $k_z$ point is measured at a different $h\nu$, the $E(k_z)$ cut carries considerably lower statistics and much sparser \textbf{k}-space sampling than the $E(k_x)$ cut, acquired as a single image at fixed $h\nu$.  

The two cuts reveal the same valence-band dispersions, including band positions, curvatures, and the deep Pb~$6s$--Se~$4s$ feature near $-8$~eV, regardless of whether $k$ is varied in-plane or out-of-plane.
This is the direct experimental signature of the bulk sensitivity of soft-X-ray ARPES at the photon energies used here ($400$--$900$~eV)~\cite{Strocov2014_SXARPES}.
We note, however, that the band dispersions $E(k_z)$ may exhibit intrinsic $k_z$ broadening arising from the limited photoelectron mean free path~\cite{Strocov2003_kz_broadening}.

Figure~\ref{fig:SI_oop_iso} presents the out-of-plane iso-energy maps at the same four binding energies as the main text.
They exhibit the same qualitative features: isolated L-pocket projections at $E - E_\Sigma = 0$, bright $\langle 100 \rangle$ ridges and $\Sigma$- and $\Gamma$-centered features at intermediate depths, and an octagonal ring near the zone boundary at $-3.2$~eV. The  agreement with the QP$GW$ contours is also similar to  the horizontal $\Gamma$XW plane. This indicates that the in-plane and out-of-plane ARPES data probe the same 3D bulk valence band~\cite{Strocov2014_SXARPES}.

\FloatBarrier

\section{In-plane iso-energy maps compared with PBE and HSE}

\begin{figure}[h]
\centering
\includegraphics[width=0.95\linewidth]{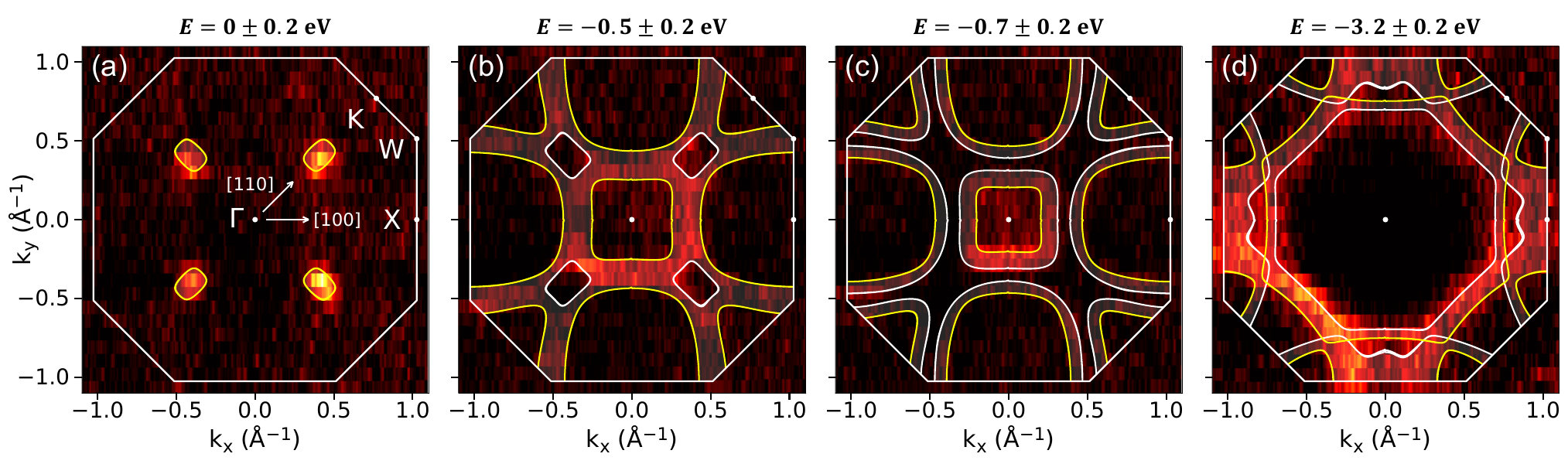}\\
\vspace{0.4em}
\includegraphics[width=0.95\linewidth]{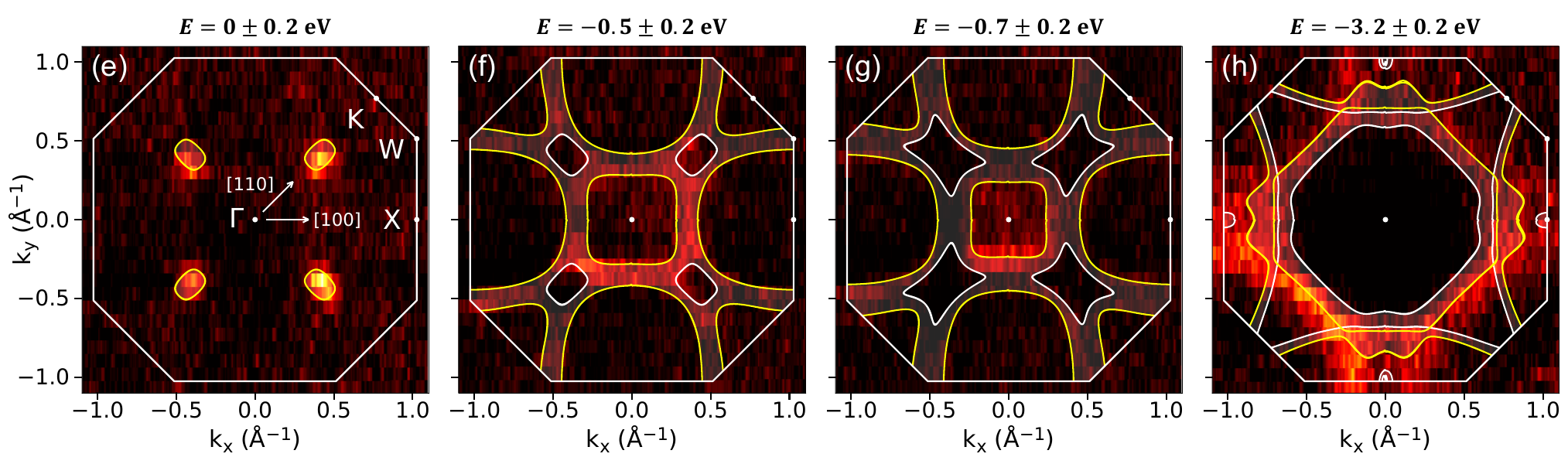}
\caption{Comparison of the PBE ((a)--(d)) and HSE ((e)--(h)) constant-energy contours (white and yellow lines, marking the top and bottom edges of the integration window) with the same ARPES iso-energy maps as Figure~5 in the main text, at $E - E_\Sigma = 0$, $-0.5$, $-0.7$, and $-3.2$~eV (left to right in each row) with $\pm 0.2$~eV integration windows.
The octagonal boundary marks the (001) surface Brillouin zone.}
\label{fig:SI_iso_pbe_hse}
\end{figure}

Figure~\ref{fig:SI_iso_pbe_hse} repeats the in-plane comparison of Figure~5 of the main text with the PBE and HSE constant-energy contours in place of the QP$GW$ ones, at the same four binding energies and with the same $\pm 0.2$~eV integration windows, referenced to each method's own $\Sigma$ maximum. 
The PBE contours at $E - E_\Sigma = 0$ and $-0.5$~eV are similar to the QP$GW$ ones and agree equally well with the ARPES maps. 
At $-0.7$~eV, however, the PBE contour at the upper edge of the integration window already radiates from the $\Sigma$ pockets to the zone boundary, whereas the measured low-intensity areas remain confined to four separate $\Sigma$ pockets. 
At $-3.2$~eV, PBE tracks the feature in the neighborhood of the X points, but its $\Gamma$-centered contour is too large, so that the ring-shaped high-intensity area extends inward past it. 
This follows from the compressed PBE band width, which moves the crossings of the $-3.0$~eV cut with the second and third valence bands outward. 
The HSE contours at $0$, $-0.5$, and $-0.7$~eV are very similar to the QP$GW$ ones and show the same good agreement with the measurements. 
At $-3.2$~eV, HSE does not capture the feature around the X points that connects to the zone boundary. It produces small closed contours around X instead of the continuous segments seen in the measurement.

\FloatBarrier

\section{Density of states}

\begin{figure}[h]
\centering
\includegraphics[width=0.8\linewidth]{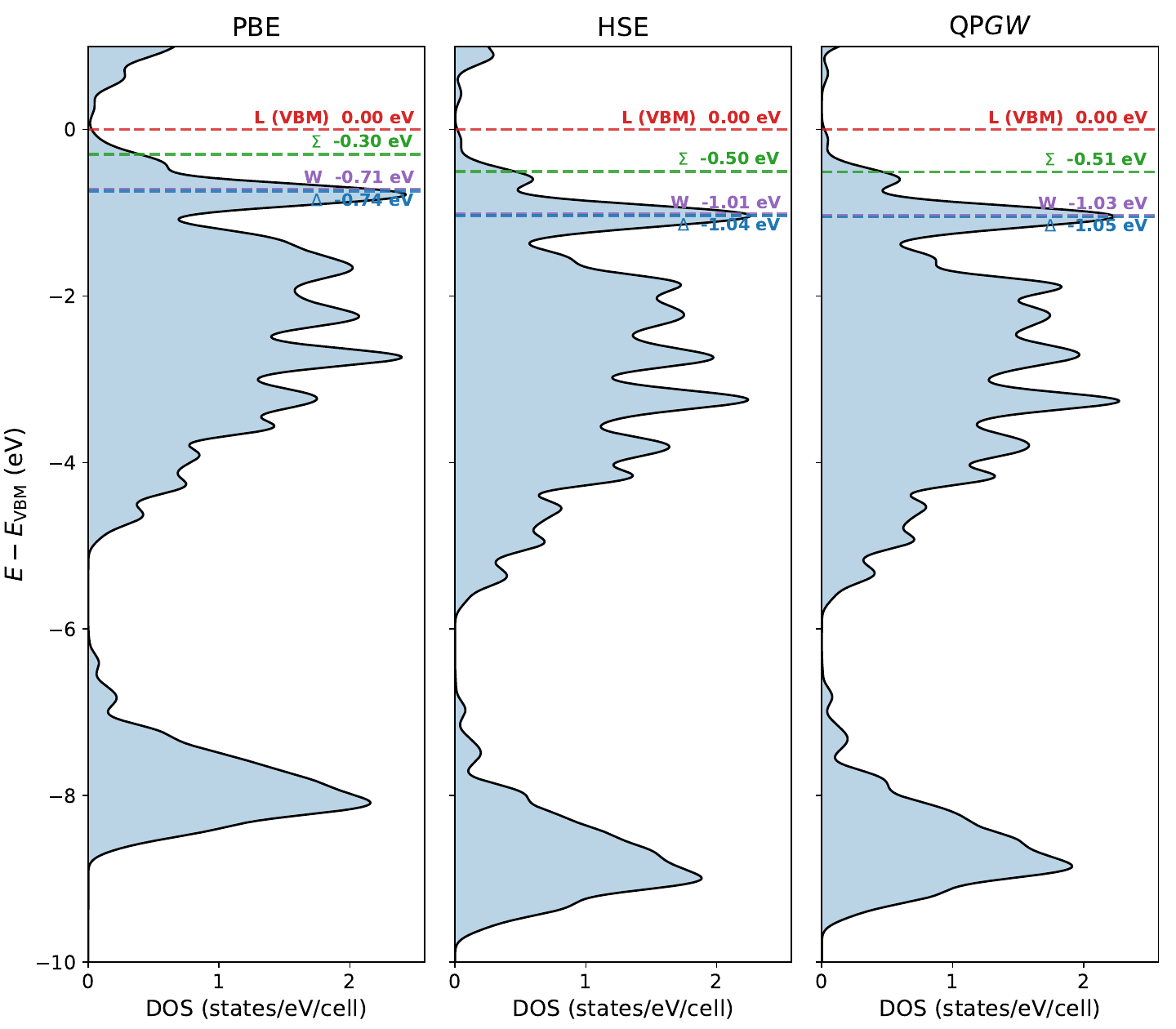}
\caption{Density of states (DOS) of PbSe computed with PBE (left), HSE (middle), and QP$GW$ (right), each referenced to its own L-point VBM.
The DOS was plotted with a Gaussian smoothing of 0.10~eV.
Horizontal dashed lines mark the L-point VBM (red), the $\Sigma$ maximum along $\Gamma$--K (green), the W-point maximum (purple), and the $\Delta$ maximum along $\Gamma$--X (blue), whose energies are given in each panel.}
\label{fig:SI_DOS}
\end{figure}

Figure~\ref{fig:SI_DOS} compares the PBE, HSE, and QP$GW$ density of states (DOS) of PbSe between $-10$ and $1$~eV.
The PBE band width is compressed relative to HSE and QP$GW$.
The Se~$4p$-dominated upper valence manifold extends down to approximately $-5.0$~eV with PBE, compared with $-5.7$~eV with HSE and QP$GW$, and the Pb~$6s$--Se~$4s$ band spans $-6.2$ to $-8.7$~eV with PBE, compared with $-6.7$ to $-9.6$~eV with HSE and $-6.5$ to $-9.5$~eV with QP$GW$.
The same compression is seen in the positions of the $\Sigma$, W, and $\Delta$ maxima marked in each panel, which lie 0.2--0.3~eV closer to the VBM with PBE.
The DOS of HSE and QP$GW$ are very similar throughout the valence band.
The only visible difference is a slightly smaller separation between the upper valence manifold and the deep Pb~$6s$--Se~$4s$ band with QP$GW$, which is about 0.15~eV closer than with HSE. 

\FloatBarrier

\section{Orbital-resolved band structures}

\begin{figure}[h]
\centering
\includegraphics[width=0.48\linewidth]{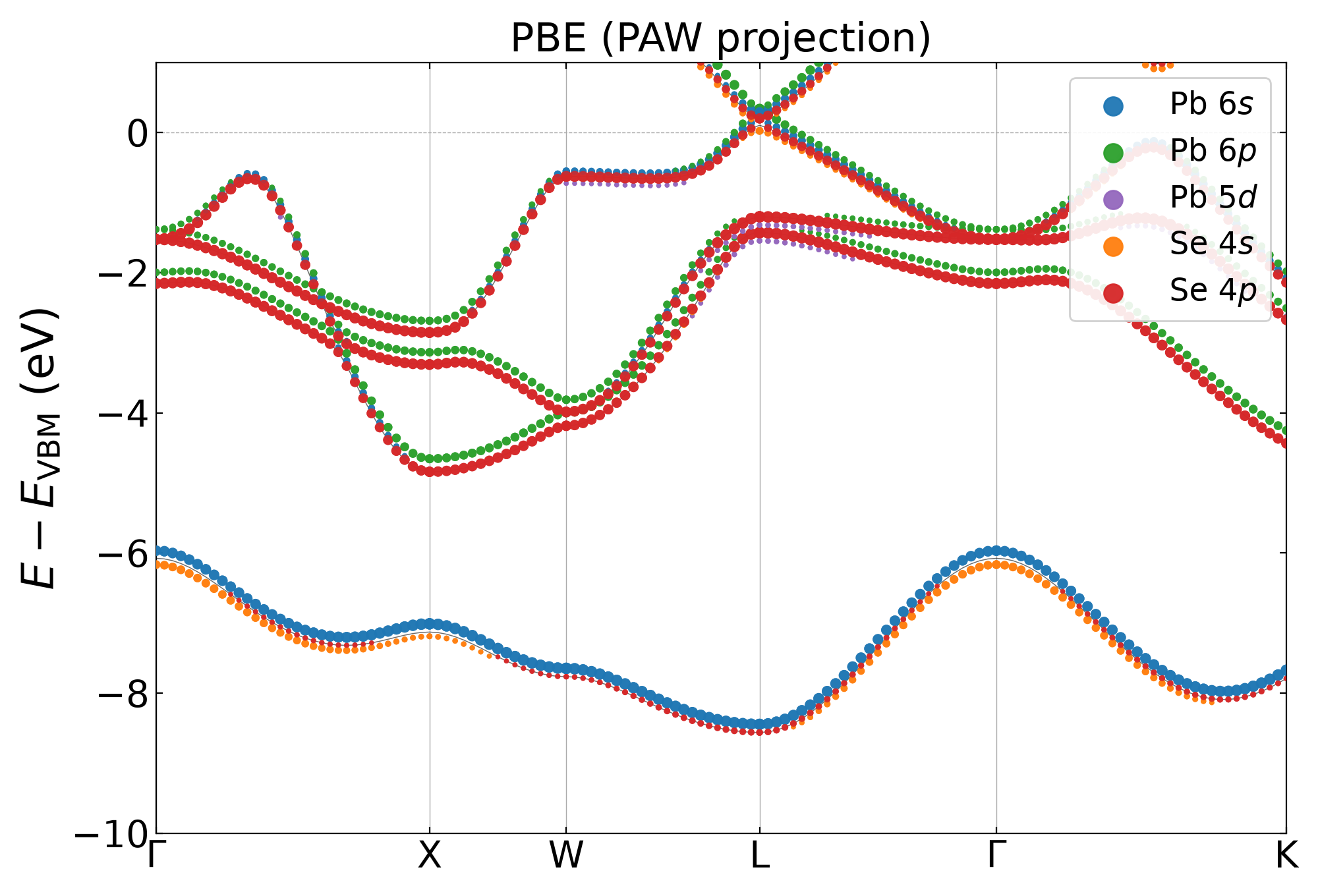}\hfill
\includegraphics[width=0.48\linewidth]{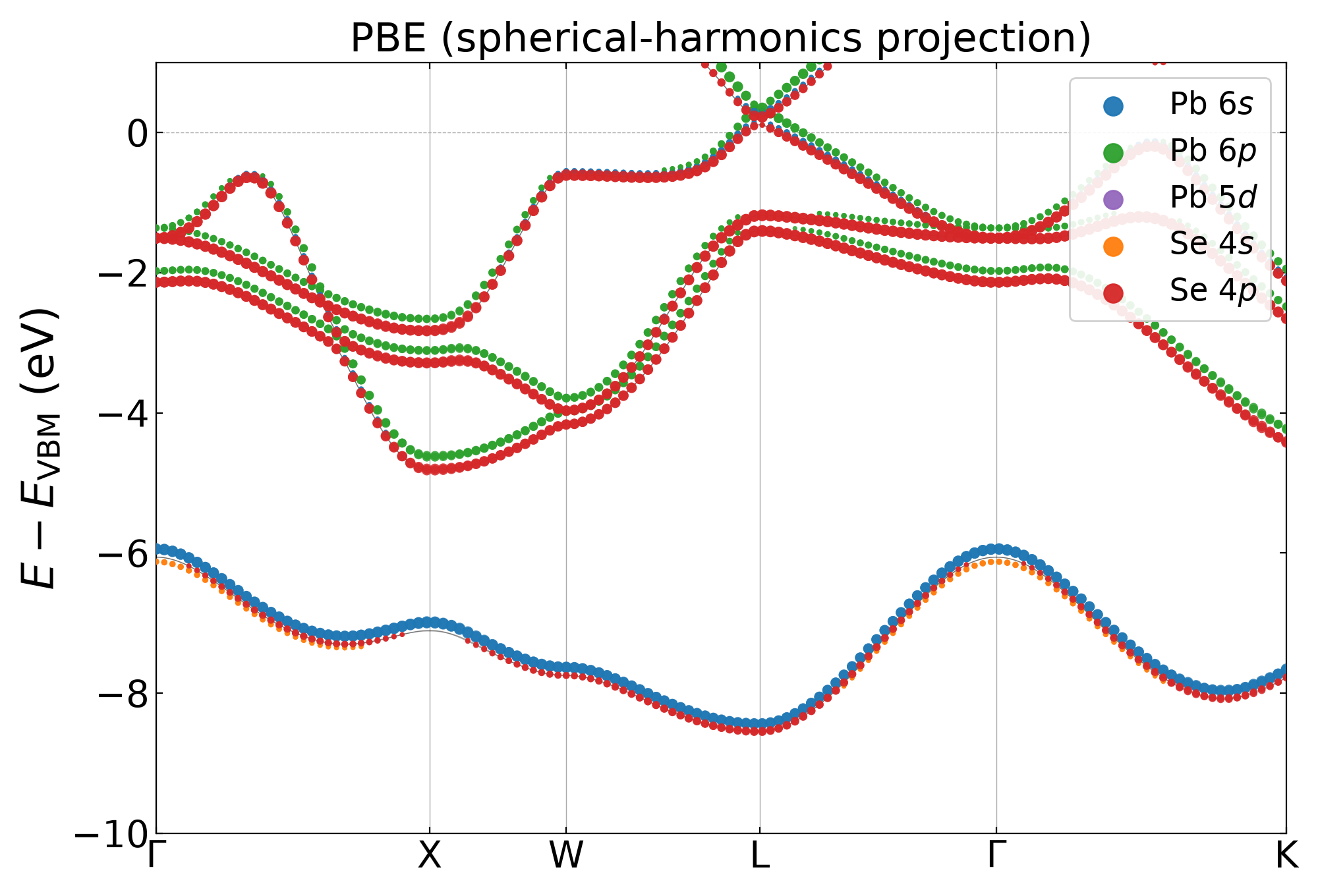}\\[0.25em]
\includegraphics[width=0.48\linewidth]{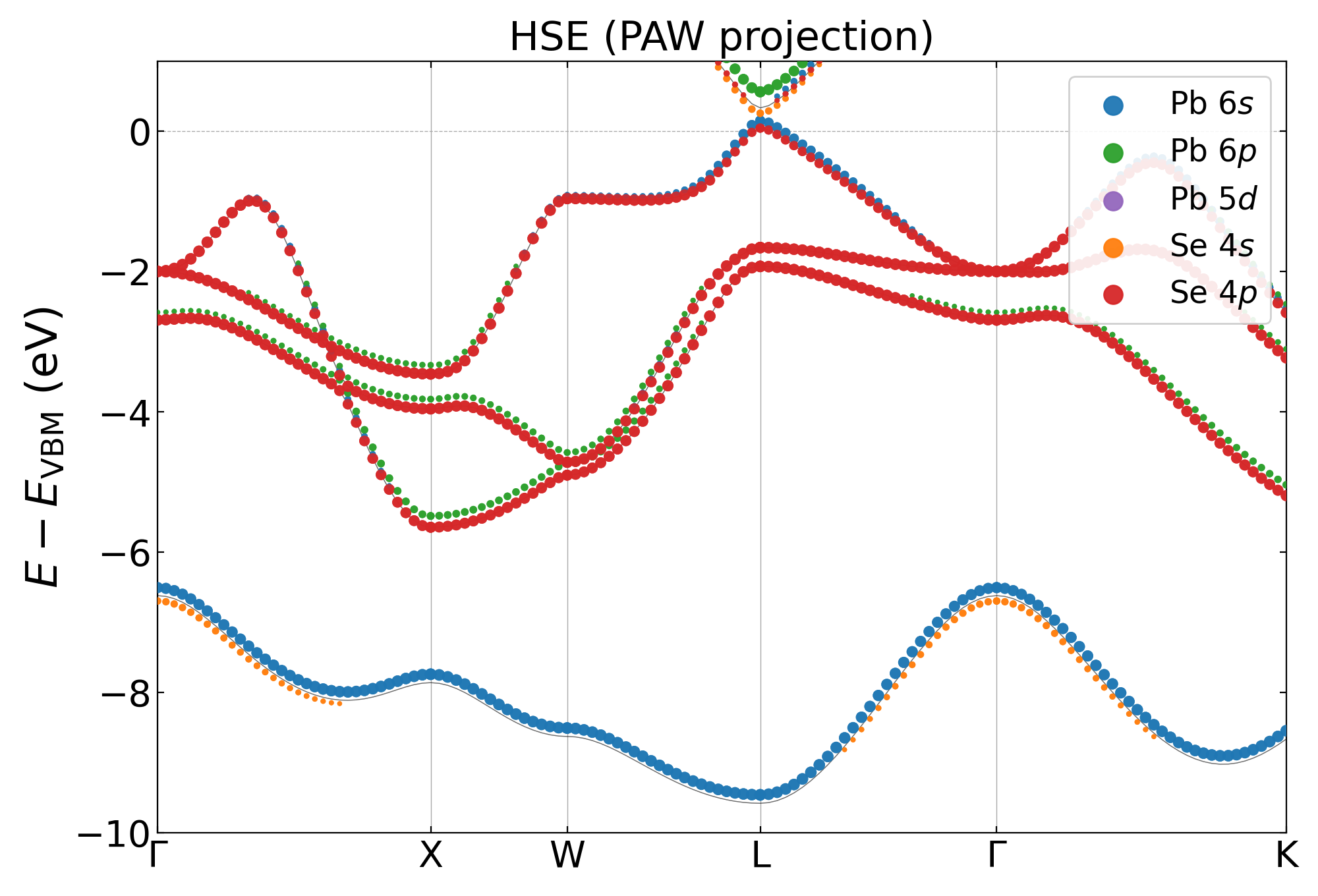}\hfill
\includegraphics[width=0.48\linewidth]{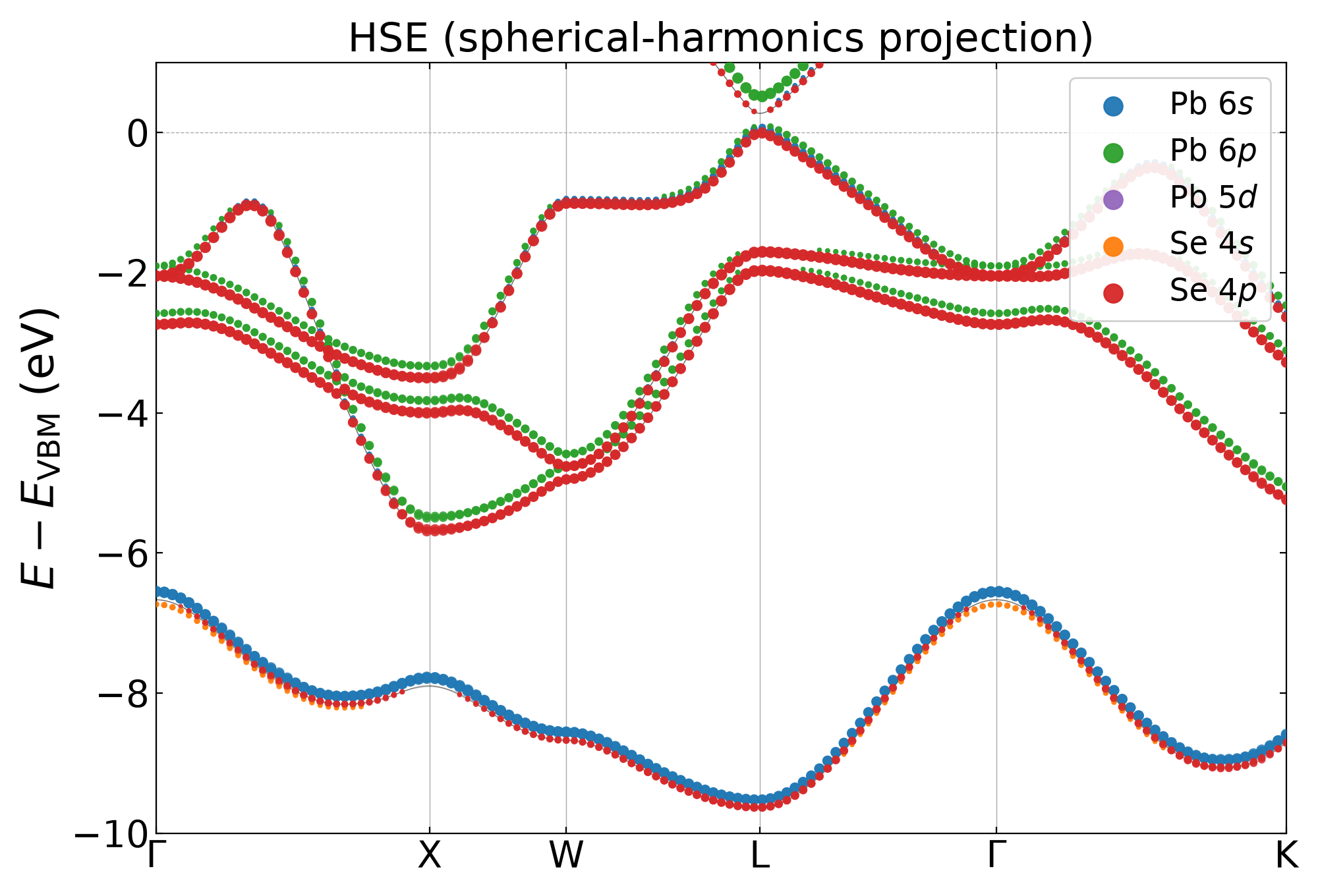}\\[0.25em]
\makebox[0.48\linewidth]{}\hfill
\includegraphics[width=0.48\linewidth]{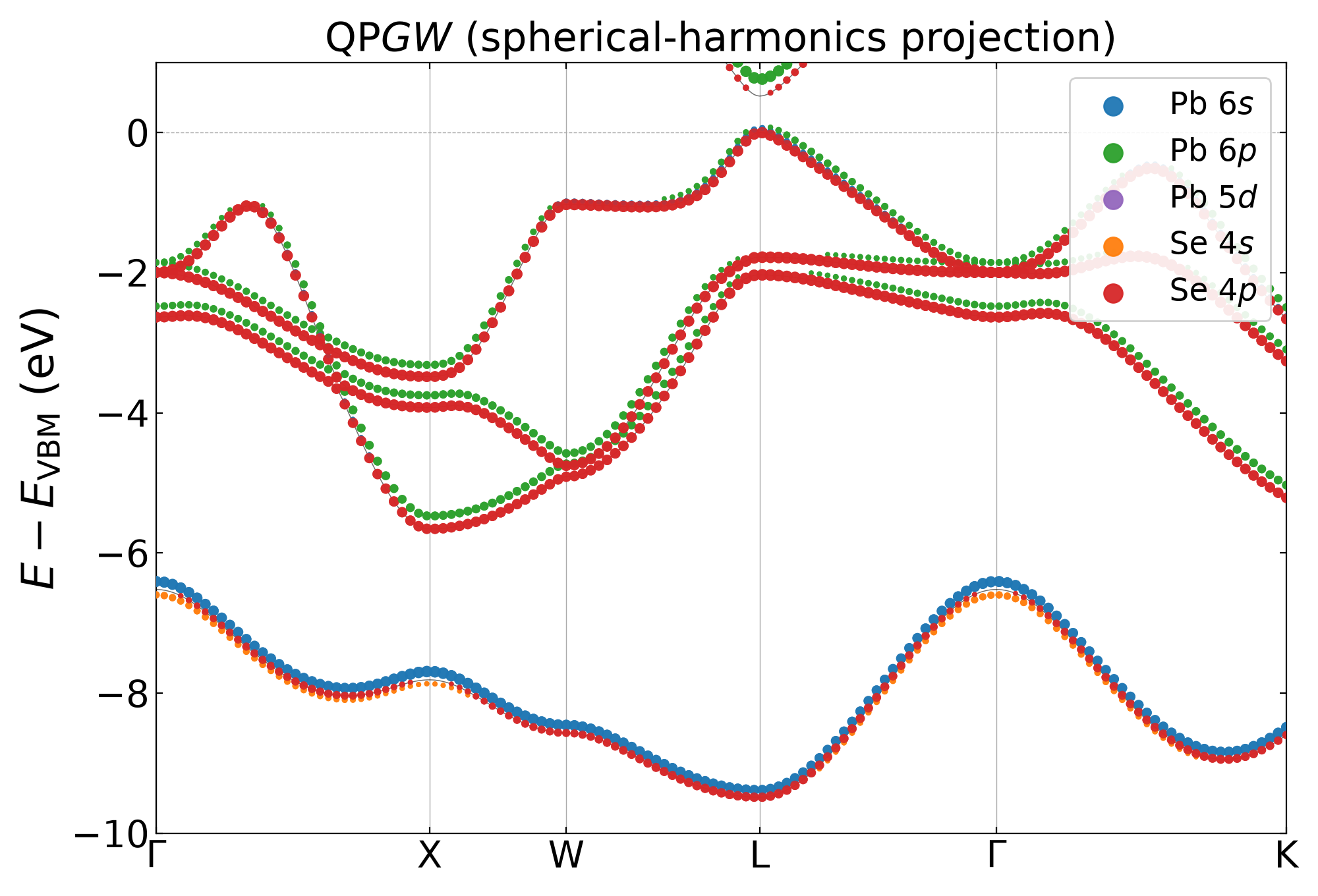}
\caption{Orbital-resolved band structure of PbSe: the PAW-projector projection (from \texttt{PROCAR}; left column, PBE and HSE) and the spherical-harmonics projection (right column, PBE, HSE, and QP$GW$).
Colors denote Pb~$6s$ (blue), Pb~$6p$ (green), Se~$4s$ (orange), Se~$4p$ (red), and Pb~$5d$ (purple).
Circle area scales with the squared orbital weight, with a small vertical offset per orbital group so that admixtures remain visible.}
\label{fig:SI_orbital}
\end{figure}

Figure~\ref{fig:SI_orbital} shows the orbital-resolved band structure of PbSe along $\Gamma$--X--W--L--$\Gamma$--K, with the PAW-projector scheme in the left column (PBE and HSE) and the spherical-harmonics projection in the right column (PBE, HSE, and QP$GW$).
Circles are drawn at each $(\text{band},\mathbf{k})$ with area proportional to the squared orbital weight, normalized so that the five group weights $\{\text{Pb}\,6s, \text{Pb}\,6p, \text{Pb}\,5d, \text{Se}\,4s, \text{Se}\,4p\}$ sum to unity; small vertical offsets are applied per orbital group so that admixtures remain visible when one character dominates.

The panels compare two projection schemes.
In the PAW-projector scheme (left column of Figure~\ref{fig:SI_orbital}), orbital weights are read directly from the VASP \texttt{PROCAR} file generated with \texttt{LORBIT}\,=\,11 (post-processed with \texttt{vaspvis}~\cite{Yu2021_vaspvis,2022_Dardzinski_BestPracticesDFTInorganicInterfacesBO_JPHYS-CONDENSMAT}).
These weights are the squared overlaps of each Bloch state with the atom-centered PAW projector functions of a given angular momentum, which is the standard and physically well-defined notion of orbital character in a PAW calculation~\cite{1999_Kresse_UltrasoftPseudopotentialsAumented_PRB}.
In the spherical-harmonics scheme (right column of Figure~\ref{fig:SI_orbital}), the weights are instead overlaps with atom-centered, $l$-resolved spherical-harmonics trial orbitals $\{\varphi_\alpha\}$ defined in \texttt{wannier90.win} (Pb $s$/$p$/$d$, Se $s$/$p$; 26 spinor functions), evaluated on the band path by a two-step Wannier-interpolated atomic projection.
First, a projection-only VASP pass (\texttt{ALGO}\,=\,\texttt{None}, \texttt{NELM}\,=\,1, \texttt{LWANNIER90}\,=\,\texttt{.TRUE.}, \texttt{LWANNIER90\_RUN}\,=\,\texttt{.FALSE.}) on the method's \texttt{WAVECAR} computes the overlap matrix $A_{n\alpha}(\mathbf{k}) = \langle \psi_{n\mathbf{k}} | \varphi_\alpha \rangle$ on the SCF $\mathbf{k}$ mesh.
Second, this overlap matrix is combined with the unitary matrix $U(\mathbf{k})$ of the standard Wannier interpolation (Section~II of the main text) into the Wannier-gauge projector $c_{i\alpha}(\mathbf{k}) = [U^{\dagger}(\mathbf{k}) A(\mathbf{k})]_{i\alpha}$, which is Fourier-interpolated onto the band-path $\mathbf{k}'$ points and contracted with the Wannier-Hamiltonian eigenvectors $v_{i\mu}(\mathbf{k}')$ to give $w_\alpha(\mu,\mathbf{k}')=|\sum_i v_{i\mu}(\mathbf{k}') c^*_{i\alpha}(\mathbf{k}')|^2$.
This preserves the band dispersions of the standard Wannier interpolation while adding the atomic-orbital character.
For QP$GW$ the spherical-harmonics route is the only one available, because a band-path \texttt{PROCAR} cannot be produced by a standard VASP non-self-consistent run on top of a QP$GW$ \texttt{WAVECAR}; for PBE and HSE both routes are available.

Although the two schemes rely on different projector functions, they assign a similar orbital character to every band along the entire path, as seen by comparing the left and right columns of Figure~\ref{fig:SI_orbital} for PBE and HSE.
For the exact magnitudes of the orbital weights, the PAW-projector weights (left column of Figure~\ref{fig:SI_orbital}) should be regarded as canonical, because their projectors are intrinsic to the electronic-structure calculation~\cite{1994_Blochl_ProjectorAugmentedWaveMethod_PRB,1999_Kresse_UltrasoftPseudopotentialsAumented_PRB}.
The comparison at fixed scheme (HSE vs.\ QP$GW$ in the right column of Figure~\ref{fig:SI_orbital}) shows HSE and QP$GW$ to be nearly indistinguishable, mirroring the close agreement of their valence dispersions in Figure~3(a) of the main text.

\FloatBarrier

\section{\label{sec:SI_transport}Temperature-dependent Seebeck coefficient}

\begin{figure}[h]
\centering
\includegraphics[width=1.0\textwidth]{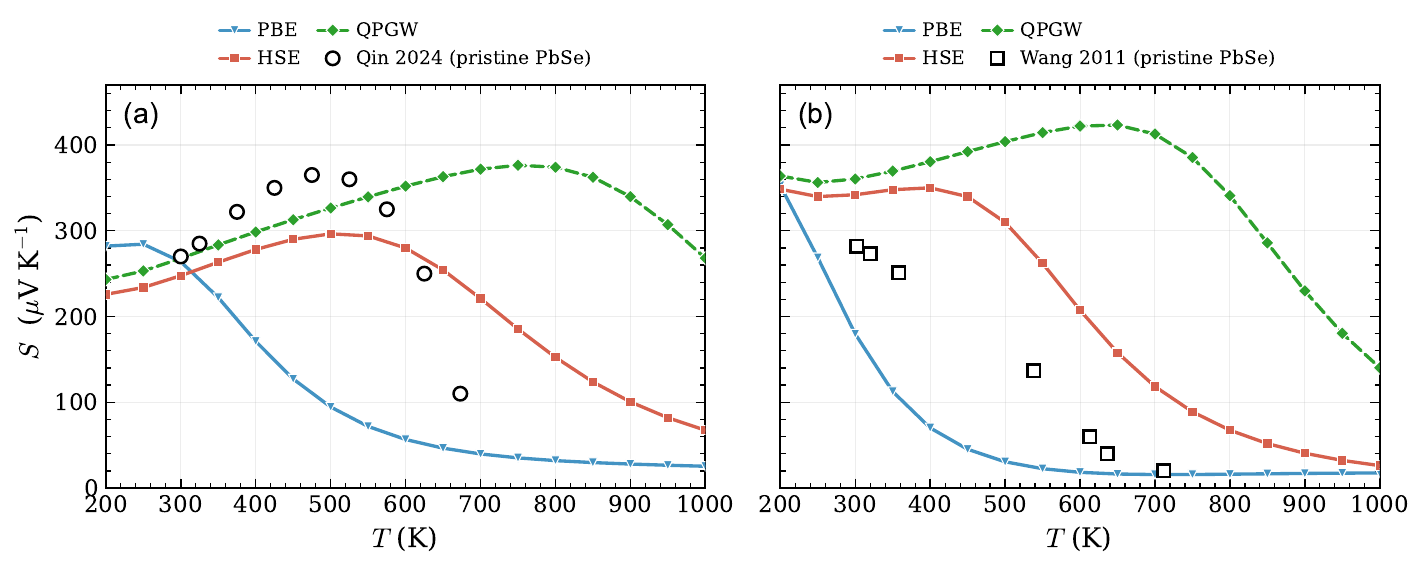}
\caption{Temperature-dependent Seebeck coefficient $S(T)$ of two pristine p-type PbSe samples at fixed acceptor density, computed from the PBE, HSE, and QP$GW$ Wannier Hamiltonians.
(a)~The pristine sample of Ref.~\cite{Qin2024_Science_gridplain}, $p_{\mathrm{H}} = 2.89\times 10^{18}$~cm$^{-3}$.
Open circles: the measured $S(T)$ of that sample, digitized from Figure~S1B of Ref.~\cite{Qin2024_Science_gridplain}.
(b)~The pristine sample of Ref.~\cite{Wang2011_PbSe_thermoelectric}, $p_{\mathrm{H}} = 1.20\times 10^{18}$~cm$^{-3}$.
Open squares: its measured $S(T)$, read from Figure~1(a) of Ref.~\cite{Wang2011_PbSe_thermoelectric}.}
\label{fig:svt}
\end{figure}

In the following, we present the temperature dependence of the Seebeck coefficient of pristine PbSe, obtained by solving the charge-neutrality condition at a fixed acceptor density equal to the measured room-temperature Hall density $p_{\mathrm{H}}$ of each sample. These calculations are the only ones that take an experimental input: $p_{\mathrm{H}} = 2.89\times 10^{18}$~cm$^{-3}$ for the pristine sample of Ref.~\cite{Qin2024_Science_gridplain} and $1.20\times 10^{18}$~cm$^{-3}$ for the pristine sample of Ref.~\cite{Wang2011_PbSe_thermoelectric}. Taking the acceptor density from a room-temperature or lower-temperature Hall measurement is justified because thermal minority carriers are negligible there~\cite{Ravich1970_LeadChalcogenides,Chasapis2015_PbSe_twoband}, and the acceptors are assumed to remain fully ionized at all temperatures, so that $p_{\mathrm{H}}$ is temperature independent. 

Figure~\ref{fig:svt}(a) compares the computed $S(T)$ with the measured data of the pristine sample of Ref.~\cite{Qin2024_Science_gridplain}.
The three methods produce different band gaps ($0.12$, $0.26$, and $0.52$~eV for PBE, HSE, and QP$GW$ vs.\ the experimental $\approx 0.27$~eV).
At 300~K, QP$GW$ gives the closest value among the methods still in the hole-dominated regime ($+268$ vs.\ the measured $+270\,\mu$V/K); the nominally similar PBE value ($+264\,\mu$V/K) is accidental, since PBE ($n/p \approx 6\%$ at 300~K) has already passed its bipolar maximum and describes the wrong transport regime.
HSE tracks the measured trend, yielding $+248\,\mu$V/K at 300~K and a bipolar maximum at 500~K, within the observed 475--525~K window. However, the peak value is $\sim 19\%$ lower than experiment.
The HSE peak-position agreement follows from its gap matching the experimental 0.27~eV, while the QP$GW$ agreement at 300~K, where the transport is still hole-dominated, indicates that QP$GW$ gives the most accurate description of the valence band structure.

Panel~(b) of Figure~\ref{fig:svt} applies the fixed-acceptor protocol to another pristine sample from Ref.~\cite{Wang2011_PbSe_thermoelectric}.
Its measured Seebeck coefficient declines monotonically with increasing temperature, starting from 300~K, i.e.\ the sample sits at or past its bipolar maximum already at room temperature.
This behavior is not reproduced by all three methods.
For HSE, the bipolar maximum at this lower doping moves down to $\approx 400$~K and becomes nearly flat ($+350\,\mu$V/K at the maximum vs.\ $+342\,\mu$V/K at 300~K).
PBE is past the bipolar maximum at this doping level, and  has fallen to $+180\,\mu$V/K by 300~K. QP$GW$ has a maximum at $\approx 650$~K due to its larger gap.

\begin{figure}[h]
\centering
\includegraphics[width=1.0\textwidth]{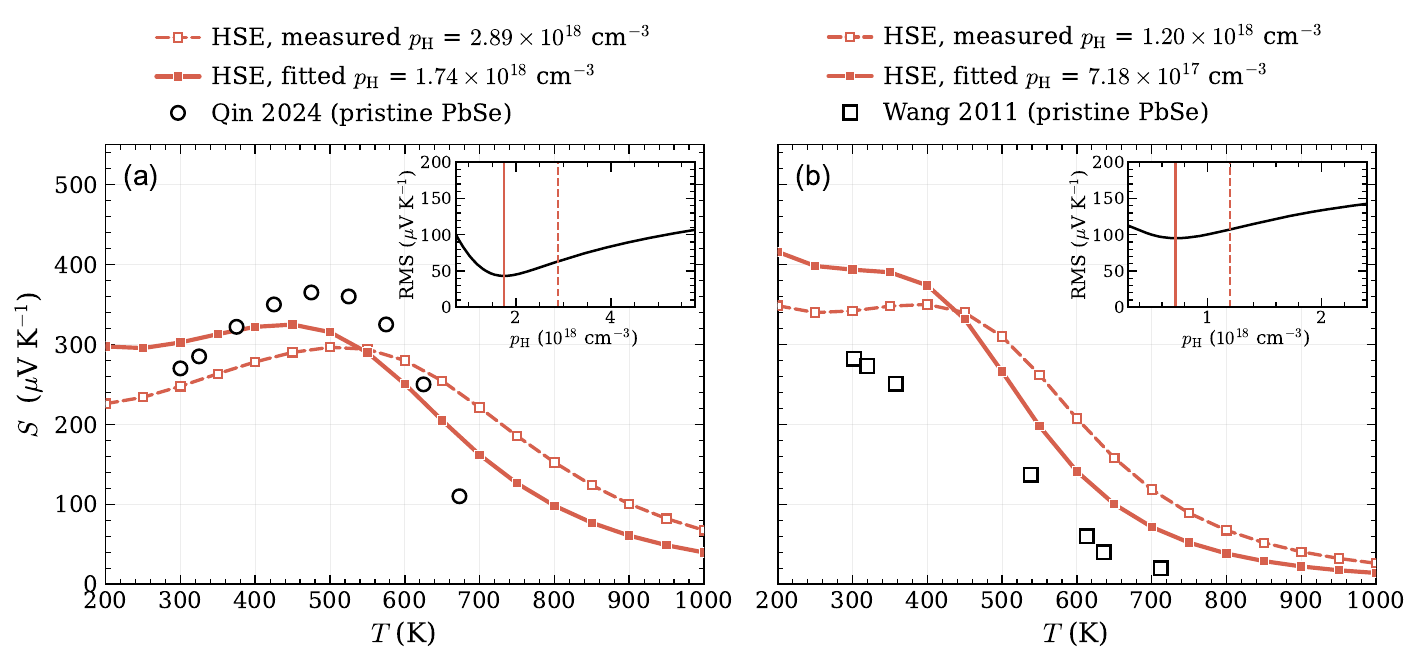}
\caption{HSE Seebeck coefficient $S(T)$ of the two pristine PbSe samples from Figure~\ref{fig:svt} with two choices of the net doping density $p_{\mathrm{H}}$.
Dashed lines: $p_{\mathrm{H}}$ equal to the measured room-temperature Hall density, as in Figure~\ref{fig:svt}.
Solid lines: $p_{\mathrm{H}}$ obtained by a least-squares fit.
Insets: root-mean-square (RMS) deviation as a function of $p_{\mathrm{H}}$, with the measured (dashed) and fitted (solid) values marked.
(a)~Sample from Ref.~\cite{Qin2024_Science_gridplain}: fitted $p_{\mathrm{H}} = 1.74\times 10^{18}$~cm$^{-3}$, RMS deviation 43~$\mu$V/K compared with 63~$\mu$V/K at the measured value.
(b)~Sample from Ref.~\cite{Wang2011_PbSe_thermoelectric}: fitted $p_{\mathrm{H}} = 7.18\times 10^{17}$~cm$^{-3}$, RMS deviation 95~$\mu$V/K compared with 107~$\mu$V/K at the measured value.}
\label{fig:svt_fit}
\end{figure}

Aside from the constant relaxation time approximation, another source of the discrepancy might be the use of the 300-K Hall density as the acceptor density for the calculation.
At such low doping, the Hall density may carry a significant electron contribution even at room temperature, so that the true acceptor density is lower and the computed curves should shift toward lower temperatures accordingly.
We tested this by treating the net doping density as a fitting parameter, scanning it on a linear grid between 0.25 and 2 times the measured value and selecting the minimum of the root-mean-square (RMS) deviation between the computed and the measured $S(T)$ at the measured temperatures (Figure~\ref{fig:svt_fit}).
The fitted doping densities are both about 40\% lower than the measured Hall densities ($1.74\times 10^{18}$ vs.\ $2.89\times 10^{18}$~cm$^{-3}$ for the sample of Ref.~\cite{Qin2024_Science_gridplain} and $7.18\times 10^{17}$ vs.\ $1.20\times 10^{18}$~cm$^{-3}$ for the sample of Ref.~\cite{Wang2011_PbSe_thermoelectric}).
With these densities, the computed Seebeck peak shifts toward lower temperatures and the HSE curves follow the experimental response more closely.

\FloatBarrier

\bibliography{reference}